\documentclass[fleqn,usenatbib]{mnras}

\usepackage{newtxtext,newtxmath}

\usepackage[T1]{fontenc}
\usepackage{ae,aecompl}

\usepackage{graphicx}	
\usepackage{amsmath}	
\usepackage{amssymb}	

\title[Rapid Mass Loss Dynamics]{A Drop in the Pond: The Effect of Rapid Mass Loss on the Dynamics and Interaction Rate of Collisionless Particles}

\author[Z. Penoyre \& Z. Haiman]{
Zephyr Penoyre,$^{1}$\thanks{E-mail: zpenoyre@astro.columbia.edu}
Zolt\'an Haiman$^{1,2}$
\\
$^{1}$Department of Astronomy, Columbia University, 550 W. 120th St., New York, NY, 10027, USA\\
$^{2}$Department of Physics, New York University, New York, NY 10003, USA
}

\date{Accepted September $21^\mathrm{st}$ 2017}

\pubyear{2017}

\begin{document}
\label{firstpage}
\pagerange{\pageref{firstpage}--\pageref{lastpage}}
\maketitle

\begin{abstract}
In symmetric gravitating systems experiencing rapid mass loss,
particle orbits change almost instantaneously, which can lead to the
development of a sharply contoured density profile, including singular caustics for collisionless systems.
This framework can be used to model a variety of dynamical systems, such as accretion disks following a massive black hole merger
 and dwarf galaxies following violent early star formation feedback. Particle interactions in the high-density peaks seem a 
 promising source of observable signatures of these mass loss events (i.e. a possible EM counterpart for black hole mergers or 
 strong gamma-ray emission from dark matter annihilation around young galaxies), because the interaction rate depends on the 
 square of the density. We study post-mass-loss density profiles, both analytic and numerical, in idealised cases and present 
 arguments and methods to extend to any general system. An analytic derivation is presented for particles on Keplerian orbits 
 responding to a drop in the central mass. We argue that this case, with initially circular orbits, gives the most
sharply contoured profile possible. We find that despite the presence
of a set of singular caustics, the total particle interaction rate is
reduced compared to the unperturbed system; this is a result of the
overall expansion of the system dominating over the steep caustics.  Finally we
argue that this result holds more generally, and the loss of central
mass \textit{decreases} the particle interaction rate in any physical
system.
\end{abstract}




\section{Introduction}

There are a variety of astrophysical systems which experience mass
loss on a time-scale much shorter than the dynamical time of the
system, leading to a significant shift in the dynamics.  

One example
of this phenomenon, highlighted in the recent literature, is the
merger of a binary black hole (BH): the burst of gravitational waves
during the last stage of the merger typically carries away a few
percent of the binary's rest-mass.  This basic prediction of general
relativity (GR) has been confirmed by LIGO observations of
stellar-mass BH mergers, which show that a significant fraction of the
BHs' total mass is lost \citep{LIGO_BBH,GW170104}.

Several studies have examined the impact that this mass-loss would
imprint on a circumbinary disk, both in the context of super-massive
\citep{Schnittman08,O'Neill09,Megevand09,
  Corrales10,Rossi10,Rosotti12} and stellar-mass
\citep{deMink17} BHs.
The key result of these studies is that a
sharply contoured density profile quickly emerges, with concentric
rings of large under- and over-densities, including shocks.  The
origin of this morphology is simple: the disk gas, which is initially
on circular orbits, instantaneously changes to eccentric orbits.  Over
time, the orbits at different radii shift out of phase, and in the
particle limit, intersect and create caustics (see
\citealt{Lippai08} and \citealt{Shields08} for a similar effect from
BH recoil, demonstrated by test-particle orbits).  The concentric
density spikes and shocks found in hydrodynamical simulations
correspond to these caustics \citep[e.g.][and the other references
  above]{Corrales10}.

In a different context, and on much larger spatial scales, dwarf
galaxies are believed to experience a similar rapid mass loss, when
early periods of rapid star formation (and associated supernova
feedback) blow out a large fraction of the gas from the nucleus.
Crucially, this mass ejection also occurs on a time-scale shorter than
the dynamical time.  \citet{Governato10} have shown that such rapid
supernova feedback can transfer energy to the surrounding dark matter
(DM). This model can be extended to repeated mass outflow and infall
events \citep{Pontzen12} to gradually move DM away from the center of
the galaxy and turn a cuspy profile into a core. These simple models
have been implemented in hydrodynamical simulations
\citep[e.g.][]{Governato12, Pontzen14,ElZant16}, which confirm this
basic result.

In the latter context, the focus has been on the overall
expansion of the DM core. However in principle if the outflow is dominated by a single large event rather than repeated energy bursts this collisionless DM particle
core could develop shells of overdensities and caustics, analogous the
those in the circumbinary disks.
While self-gravity will reduce the effect, these systems are of particular
interest for indirect DM detection. As suggested, e.g., in
\citet{Lake90}, they are excellent candidates for seeing $\gamma$-rays
from DM annihilation, due to their abundance in the nearby universe,
their high mass-to-light ratio, and their lack of other $\gamma$-ray
sources.  While a detection remains elusive, dwarf galaxies have been
used to put strong limits on the mass and interaction cross section of
DM particles \citep[e.g.][]{Abazajain12,Geringer15,Gaskins16}.

While the overall effect of rapid mass loss is a decrease in density
of the DM core, the presence of strong density spikes could
significantly boost the DM annihilation rate, even if these spikes
contain only a small fraction of the mass (note that the annihilation
rate is proportional to the square of the density).  This would imply
a larger $\gamma$-ray flux, and strenghten the existing limits on DM
properties.

Motivated by the above, in this paper, we compute the density profiles
of spherical, collisionless systems, following an instantaneous
mass-loss at the center.  Our models can include self-gravity, and directly
resolve the caustic structures.  We emphasize that our results are
generic, and are applicable to any quasi-spherical collisionless
system on any scale.  Our result is negative and completely
general: we find that the overall density decrease dominates over the
presence of caustics, even in the most idealized systems. As a result,
we conclude that mass-loss always decreases the net interaction
rate.\footnote{A similar result has been found in the context of
  circumbinary BH disks, where several authors have computed the
  Bremsstrahlung luminosity and found it to be lower than in the
  pre-merger disk \citep{O'Neill09,Megevand09,Corrales10}; see further
  discussion in \S~\ref{sec:conclude} below.}

This paper is organized as follows. We first present an analytic
derivation of the density profile due to mass loss in the idealized
case of a Keplerian potential (\S~\ref{Kepler}); we use these profiles
to show that there is an overall drop in the interaction rate.  We
then show that the inclusion of more realistic physical effects
generally leads to less sharply contoured profiles, thus making the
Keplerian case the upper limit for the resulting interaction rates
(\S~\ref{Other}). Finally, we present general arguments about the
interaction rate in systems undergoing simple transformations of the
density profile, and argue that these likewise imply a generic drop in
the rate, as for a Keplerian potential (\S~\ref{Boost}). We end by
summarizing the implications of this work and offering our conclusions
(\S~\ref{sec:conclude}).

\section{Circular orbits in a Keplerian potential}
\label{Kepler}

We start with one of the simplest cases possible: an initially
circular orbit of a test particle around a point mass, which at some
point instantly loses a fraction of its mass.  We choose this case not
just for its intuitive behaviour, but because it is relatively simple
to extend it to more realistic situations, and it provides an upper
limit on the interaction rate in response to rapid mass loss (as we
argue in \S~\ref{Other} below).

\subsection{Basic Keplerian orbit}
\label{BasicKepler}
Particles initially on circular orbits, if they remain bound after the
mass loss, will move on ellipses, so we start by recapitulating some
basic results for orbits in a Keplerian potential. We will utilize the
following results (derivations can be found in textbooks such as
\citealt{Binney08}).

In a Keplerian potential around a point with mass $M$ at some radius $r$,
\begin{equation}
\label{kepler_pot}
\Phi(r)=\frac{GM}{r},
\end{equation}
the radius of a particle's orbit follows
\begin{equation}
\label{kepler_basic}
r(\phi)=\frac{a (1-e^2)}{1+e \cos(\phi - \phi_0)}
\end{equation}
where $\phi$ is the phase, $\phi_0$ the initial phase, $a$ the
semi-major axis and $e$ the eccentricity.  The particle's specific
angular momentum and energy are
\begin{equation}
\label{kepler_l}
l=r v_t=r^2 \dot{\phi} = \sqrt{G M a(1-e^2)}
\end{equation}
and 
\begin{equation}
\label{kepler_E}
e=\frac{1}{2}(v_r^2 + v_t^2) - \Phi(r),
\end{equation}
both of which are conserved over an orbit. A less easily
visualised, but useful parameterisation in terms of the
eccentric anomaly $\eta$,
\begin{equation}
\label{eta}
\sqrt{1-e}\tan{\left( \frac{\phi-\phi_0}{2} \right)}=\sqrt{1+e}\tan{\left( \frac{\eta}{2} \right)},
\end{equation}
allows us to express the radius more simply as
\begin{equation}
\label{kepler_r}
r(\eta)=a(1-e \cos{\eta}).
\end{equation}
The expression for the time as a function of $\eta$ can be obtained by
integrating $\dot{\phi}$ from equation~\ref{kepler_l} and using $r$ in
the form of equation~\ref{kepler_r}. This yields,
\begin{equation}
\label{kepler_t}
t(\eta)-t_0=\sqrt{\frac{a^3}{GM}}(\eta - e \sin{\eta}),
\end{equation}
where $t_0$ is the time of the first pericenter passage (i.e. if the
particle is initially at some $\eta_0$,
$t_0=\sqrt{\frac{a^3}{GM}}[\eta_0 - e \sin{\eta_0}]$).  An orbit has
been completed when $\eta=2\pi$, and hence the orbital period is
easily confirmed from equation~\ref{kepler_t} to be
\begin{equation}
T=2\pi\sqrt{\frac{a^3}{GM}}.
\end{equation}

\subsection{Response to mass loss}
\label{KeplerResponse}

\begin{figure}
\includegraphics[width=\columnwidth]{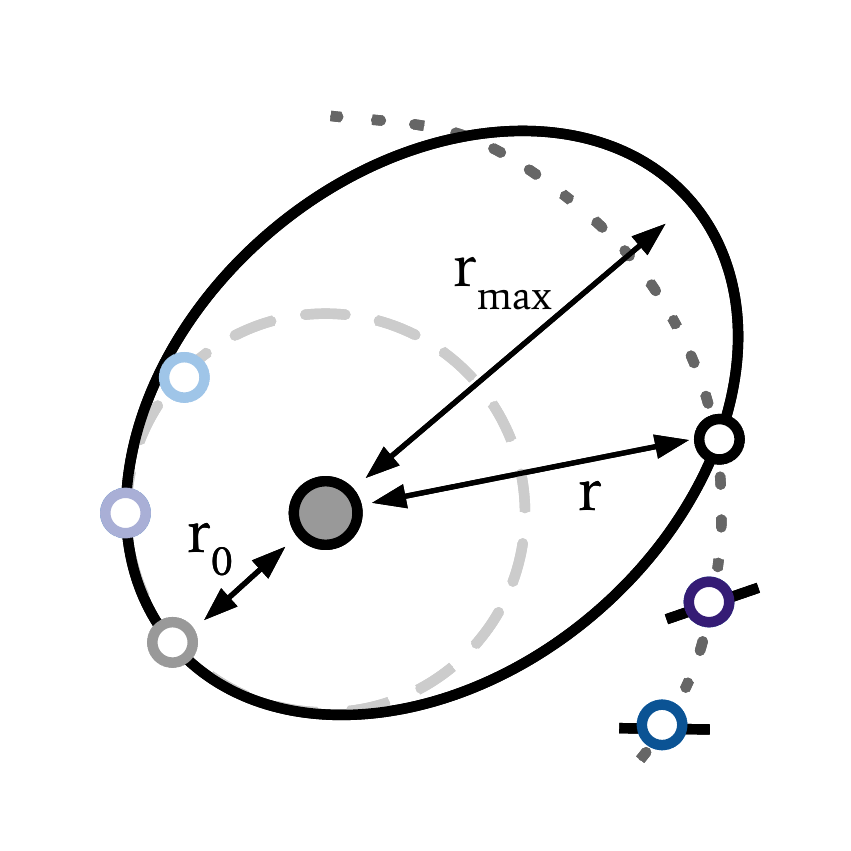}
\caption{A particle is initially on a circular orbit with radius $r_0$
  (grey dashed) around a central mass (large dark circle). At some
  time there is an instant drop in the central mass; the position of
  the particle at that time is shown as a light grey circle. The
  subsequent elliptical orbit is shown as the black solid ellipse with
  apocenter $r_{\rm max}$. The location of the particle at the current
  time, at distance $r$ ($r_0<r<r_{\rm max}$), is marked by a black
  circle. Two other particles on the same initial circular orbit but
  at different initial azimuthal angles are shown (purple and blue
  circles). Their current positions are shown in full colour with a
  segment of their trajectory, while their initial positions are shown
  in lighter colours. The dotted dark grey arc shows the common
  current distance of all three particles along a circle with radius
  $r>r_0$.}
\label{ellipse}
\end{figure}

\begin{figure}
\includegraphics[width=\columnwidth]{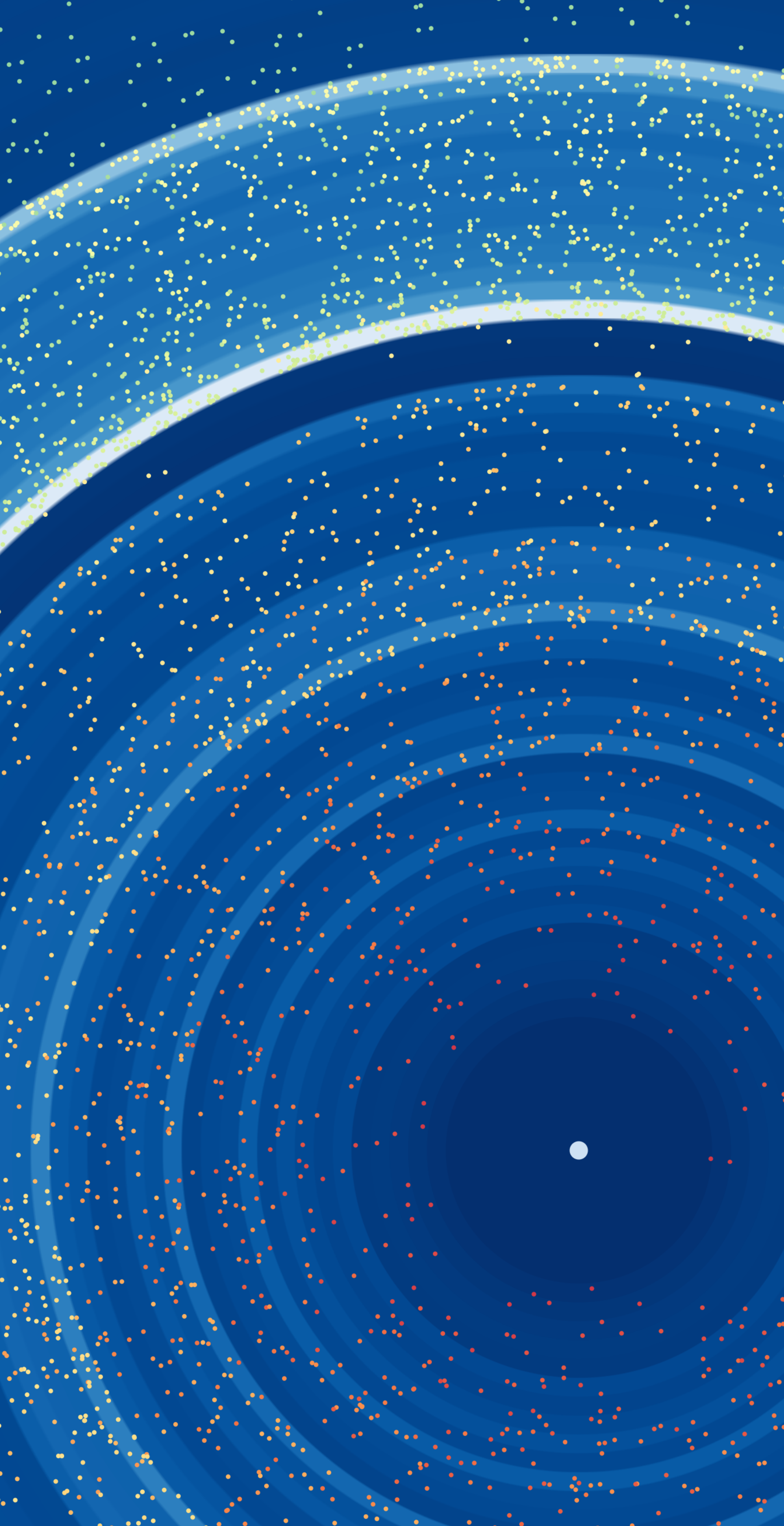}
\caption{A snapshot from a simple 2D interactive simulation of
  instantaneous mass loss in a Keplerian potential (which can be found
  here:
  \url{http://user.astro.columbia.edu/~zpenoyre/causticsWeb.html}.)
  Each dot is a particle, initially on a circular orbit and coloured
  by initial radius. The background colour shows the number density of
  particles in that circular shell, with lighter corresponding to
  higher density. The grey circle shows the point mass, whose mass has
  been dropped by 20\%, causing every particle to continue on a new
  elliptical orbit. Where many orbits overlap circular overdensities
  develop and move outwards.}
\label{web}
\end{figure}

To obtain the evolution of a spherical system, we start with a single
particle on a circular orbit, initially at some radius
$r_0$. Figure~\ref{ellipse} shows an illustration and our notation.
In the circular case, $e=0$ and hence $a=r_0$. (The corresponding
solution for non-circular orbits is given in
Appendix~\ref{EllipseKepler}.)  When the central mass instantly drops
from $M$ to $m<M$, the particle's orbit is instantly changed. The
particle is now less tightly bound, and has been given a boost in
energy (i.e. a less negative gravitational potential) and will
continue on an elliptical orbit. The angular momentum is unchanged, as
we have given it no tangential impulse, and the position and velocity
must be conserved over the instant of mass loss.  An interactive
demonstration can be found at
\url{http://user.astro.columbia.edu/~zpenoyre/causticsWeb.html} (a
still image of which is shown in Figure~\ref{web}).

The new orbit must also be Keplerian, of the form in equation
\ref{kepler_r}. Let the new eccentricity, semi-major axis and phase be
$\epsilon$, $\alpha$ and $\psi$ respectively, and let the moment of
mass loss be $t=0$.  Since the velocity at $t=0$ is purely tangential,
the particle must be at its periapsis, and hence $\psi_0$, $\eta_0$
and $t_0$ must all be equal to 0.  Conserving angular momentum
throughout, and energy for $t>0$, we can find the properties of the new
orbit.

First let us define the dimensionless constant
\begin{equation}
\label{mu}
\mu=\dfrac{1}{2-\frac{M}{m}}.
\end{equation}
We have $r_{\rm min}=r_0$ and we find that the apoapsis is at
\begin{equation}
\label{r_max}
r_{\rm max}=\dfrac{r_0}{2\frac{M}{m} -1} = \frac{M}{m}\mu \ r_0,
\end{equation}
where $\alpha=\mu \ r_0$ and $\epsilon=1-\frac{1}{\mu}$.  Two
consequences of these results are worth noting:

\begin{itemize}
\item The physical scale of an orbit depends linearly on the initial
  radius, and the eccentricity is constant for all orbits. This means
  the orbit of any two particles with different initial radii are
  similar, differing only in their period.
\item The above solution breaks down for $m \leq \frac{M}{2}$; this
  corresponds to the particle becoming unbound and elliptical orbits
  no longer existing.
\end{itemize}

Thus for a single particle initially at $r_0$,
\begin{equation}
\label{new_r}
r(r_0,\eta)=\mu r_0 (1-\epsilon \cos{\eta})
\end{equation}
and
\begin{equation}
\label{new_t}
t(r_0,\eta)=r_0^\frac{3}{2}\sqrt{\frac{\mu^3}{Gm}}(\eta - \epsilon \sin{\eta}).
\end{equation}
While this equation does not directly yield $r$ as a function of $r_0$
and $t$, we can solve it to find $\eta=\eta(r_0,t)$ and hence find $r$
from equation~\ref{kepler_r}.

The radius of a particle at some time $t$ depends only on the initial
radius $r_0$. Hence a family of particles that start at a given
$r_0$, regardless of orbital inclination, will always be at the same
radius at any moment in time. This is illustrated in Figure
\ref{ellipse} by the three particles on the dotted circular arc,
which, while they are on different orbits, all coincide at the same radius.  As a
result, the radial motion of a spherical system, composed of
individual particles, can instead be described as that of a series of
concentric spherical shells (or cylindrical shells in 2D disks and
other axisymmetric systems).

Henceforth we will refer not to individual particles but to spherical
shells, with initial and current radii $r_0$ and $r$, which obey the
above equations.

\subsection{Recovering the density profile}
\label{KeplerDensity}

\begin{figure}
\includegraphics[width=\columnwidth]{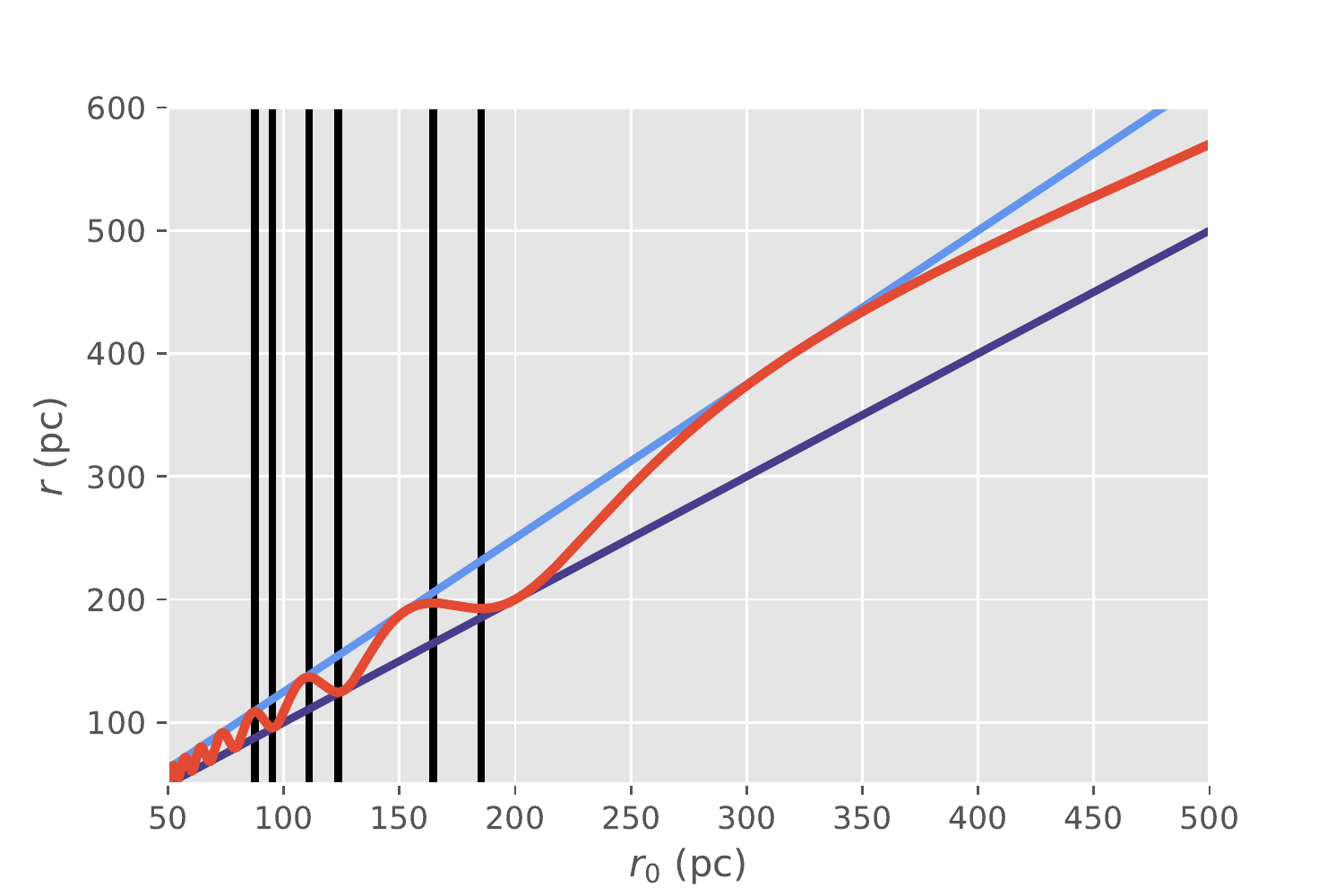}
\caption{The radius $r$ at which a particle (or spherical shell)
  resides at time $t$, as a function of its initial radius $r_0$ at
  the moment $t_0<t$ of mass loss (red curve). The pericenter of the
  elliptical orbit of each particle is $r_0$, and its apocenter is
  taken from equation~\ref{r_max} (black and blue lines,
  respectively). The outermost six turning points of the function are
  also marked by vertical black lines. Here we use an initial point
  mass of $10^{9} M_\odot$, which drops by $10\%$. The particle
  positions are shown $t=10$ Myr after the mass loss, although as discussed later the shape of the profile is self-similar and can be expressed by this curve at all times by rescaling using equation \ref{scale}.}
\label{r_r0}
\end{figure}

\begin{figure}
\includegraphics[width=\columnwidth]{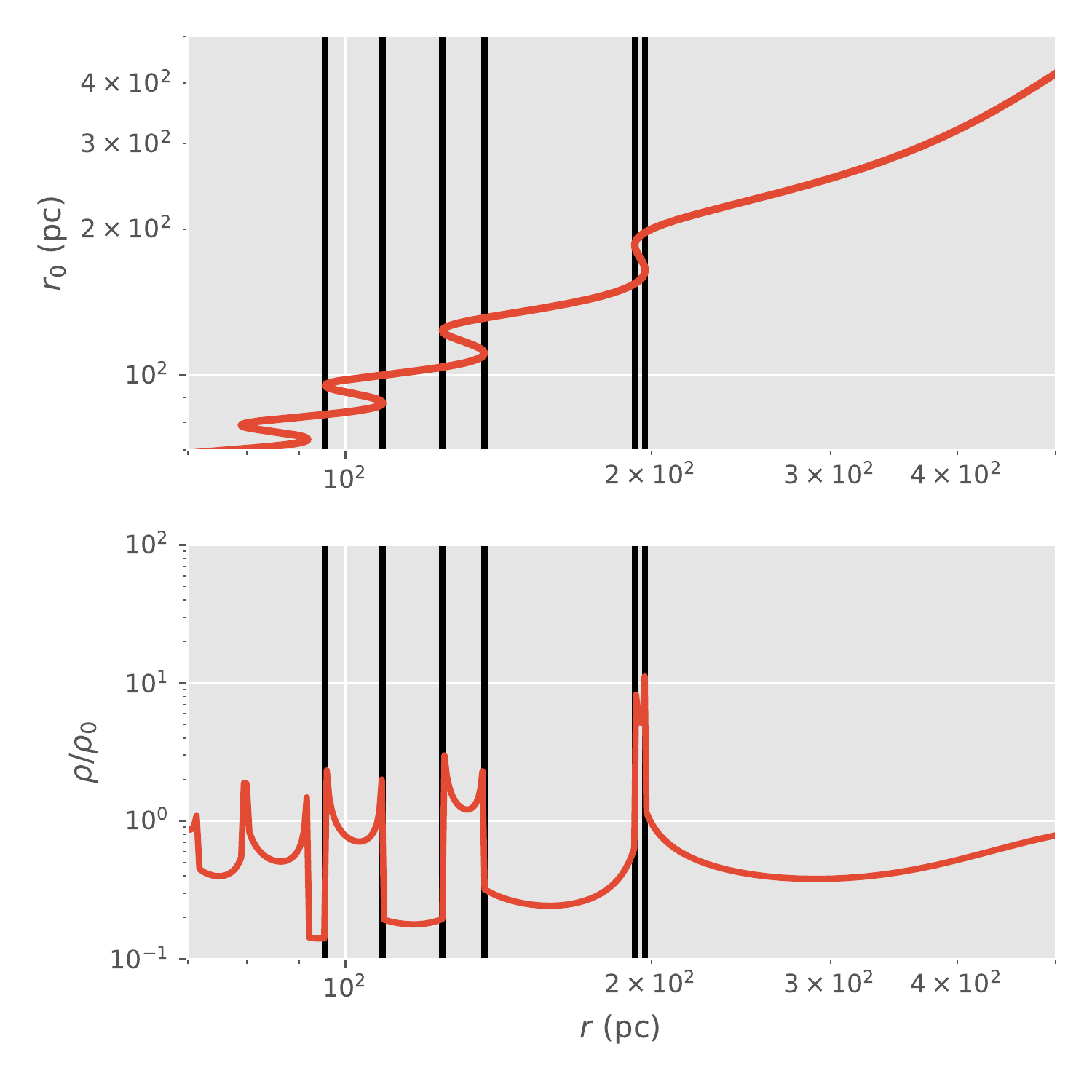}
\caption{Upper panel: The same curve as Figure \ref{r_r0}, but with
  the axes reversed.  Lower panel: The corresponding density
  profile. The two plots have a shared horizontal axis. The profile is
  analytic though computationally limited to some numerical resolution
  (detailed in Appendix \ref{KeplerSolve}) causing truly singular
  caustics to appear finite. The vertical lines show the $r$ values of
  the same turning points as shown in Figure \ref{r_r0}. Again, the shape of both profiles does not change with time and all time evolution can be captured by rescaling the radial co-ordinate using equation \ref{scale}.}
\label{r0_rho_r}
\end{figure}

With particles suddenly on a range of eccentric orbits, shells can
pass through one another and overlap, leading to overdense regions
where shells bunch up together and underdense regions where shells are
widely spread.  Our goal is to compute the time-evolution of the
density profile (and use it to compute the particle interaction
rate). The density at radius $r$ at time $t$ can be related to the initial
density profile and shell positions, using the 1D Jacobian determinant
\begin{equation}
\label{rho}
\rho(r,t)=\sum_{r_i(r_0,t)=r} \left\vert \frac{dV_0}{dV} \right\vert_{r_i} \rho_0(r_{0,i}).
\end{equation}
Here the sum is over all individual shells $i$ that are currently at a
radius $r$, but may have had different inital radii $r_{0,i}$. A
similar calculation was used in \citet{Schnittman08} under the
approximation of epicyclic orbits, whearas here we make no such
approximations.  Each shell contributes a density equal to its initial
density $\rho_0(r_0)$, multiplied by its change in infinitesimal
volume, $dV$.  For each individual shell with $dV=4 \pi r^2 dr$, we
have
\begin{equation}
\left\vert \frac{dV_0}{dV} \right\vert = \frac{r_0^2}{r^2} \left\vert \frac{dr_0}{dr} \right\vert.
\end{equation}
(Note that all analysis presented here can be easily modified to an
axisymmetric system, by replacing densities with surface densities and
the volume element with $dA=2 \pi r dr$.)  Finding the density profile
then amounts to identifying the set of shells that are at a particular
radius $r$ at a time $t$.

To simplify operations involving equations \ref{new_r} and
\ref{new_t}, we can rearrange these to make use of the fact that we
only want to recover density profiles at fixed values of $t$. From
equation~\ref{new_t}, we find $r_0=r_0(t)$ as
\begin{equation}
\label{new_r0}
r_0 = \frac{1}{\mu} \left[ \frac{G m t^2}{(\eta - \epsilon \sin{\eta})^2}  \right]^\frac{1}{3}
\end{equation}
and substituting this into equation~\ref{new_r} we obtain
\begin{equation}
\label{new_reta}
r=\left[ G m t^2 \frac{(1-\epsilon \cos{\eta})^3}{(\eta - \epsilon \sin{\eta})^2} \right]^\frac{1}{3}.
\end{equation}
Thus we have a parametric equation for $r=r(r_0,t)$.\footnote{Note
  that this solution is a generalization of the parametric solution
  for spherical collapse in cosmology -- the latter corresponds to the
  limiting case of $\epsilon=1$, i.e. pure radial orbits, in
  eq.~\ref{new_reta}.}  This solution is shown in Figure~\ref{r_r0},
for a system of initially circular orbits around a point mass of $10^9
M_\odot$, $t=10$ Myrs after the moment of a drop in the central mass
by $10\%$. The same parameters are used throughout the rest of this
section.

To interpret the result shown in Figure~\ref{r_r0} in an intuitive
way, consider the period of each shell, $T^2 \propto a^3 \propto
r_0^3$. At larger initial radii, the period becomes longer and
longer. In the extreme case, there are some particles for whom $t \ll
T$ which have barely moved from their periapsis. Further in, we see
particles for whom $t=\frac{T}{2}$, just reaching aposapsis for the
fist time. In Figure \ref{r_r0}, particles which started at roughly
200 parsecs from the central mass are just completing their first
orbit, i.e. $t=T$.  This is the outermost minimum in this figure; each
successive minimum, toward smaller radii, corresponds to those
particles which have completed another full orbit at time $t$.

Figure~\ref{r_r0} also shows that there are various locations where
multiple shells with different initial radii coincide at the same
$r$. To make this clearer, top panel of Figure \ref{r0_rho_r} shows
the same plot with the axes reversed, such that it becomes clearer to
see the radii at which multiple shells overlap. The bottom panel of
Figure \ref{r0_rho_r} shows the corresponding density profile, which
we explore next. (A more detailed discussion of how this profile is
computed can be found in Appendix~\ref{KeplerSolve}.)

The density profile in Figure \ref{r_r0} has two distinct features,
which can be understood from equation~\ref{rho}. First, large
step-like over- and underdensities, the cause of which we have already
identified as the overlap of shells from various initial radii,
i.e. they stem from the summation in equation~\ref{rho}.  Second, the
sharp density spikes (caustics) arise when the derivative
$\frac{dr}{dr_0}$ in the $\frac{dV_0}{dV}$ term goes to zero.  From
equations \ref{new_r0} and \ref{new_reta}, this derivative can be
written as
\begin{equation}
\label{dr_dr0}
\frac{dr}{dr_0}=\mu \left\{ 1-\epsilon \left[ \cos{\eta} +
\frac{3}{2}\sin{\eta} \left( \frac{\eta - \epsilon \sin{\eta}}{1 -
  \epsilon \cos{\eta}} \right) \right] \right\}.
\end{equation}
In Figure~\ref{r_r0}, the vertical lines mark the points where
$\frac{dr}{dr_0}\rightarrow 0$.  The same points are marked by
vertical lines in the top panel of Figure \ref{r0_rho_r}; their
locations clearly coincidence with the caustics in the bottom panel,
where $\rho(r)\rightarrow\infty$.  At these turning points we have a
shell whose two edges are crossing itself, i.e. where one edge of the
shell has passed through a turning point and meets the other, still to
halt, travelling in the other direction. Hence all the mass contained
within the volume element is now in enclosed in a volume that goes to
0, and the corresponding density is infinite. This can be seen clearly
in Figure \ref{r0_rho_r}, where the turning points in $r_0$ vs $r$
(which now, with the reversed axes, are vertical with a gradient going
to infinity) are again highlighted with vertical bars that correspond
perfectly to the caustics in the density profile.

Note that $r$ and $r_0$ both scale with time as $\propto
t^{2/3}$. As a result, the solutions are self-similar, and
depend only on $\frac{m}{M}$. The density profile, in particular, has
a fixed shape -- any features, such as a caustic at position $r_c$,
correspond to fixed values of $\eta$ and hence obey
\begin{equation}
\label{scale}
r_c = g(\eta) \ (Gmt^2)^\frac{1}{3},
\end{equation}
where $g(\eta)$ is a constant function of $\eta$ and is of order
$\eta^{-2/3}$. Features move outward at the ``pattern speed''
$\dot{r}_c \propto t^{-1/3}$.  Each caustic, and indeed the whole
profile, moves outward initially very fast and then slows at later
times.

\subsection{The nature of caustics}
\label{caustics}

\begin{figure*}
\includegraphics[width=\textwidth]{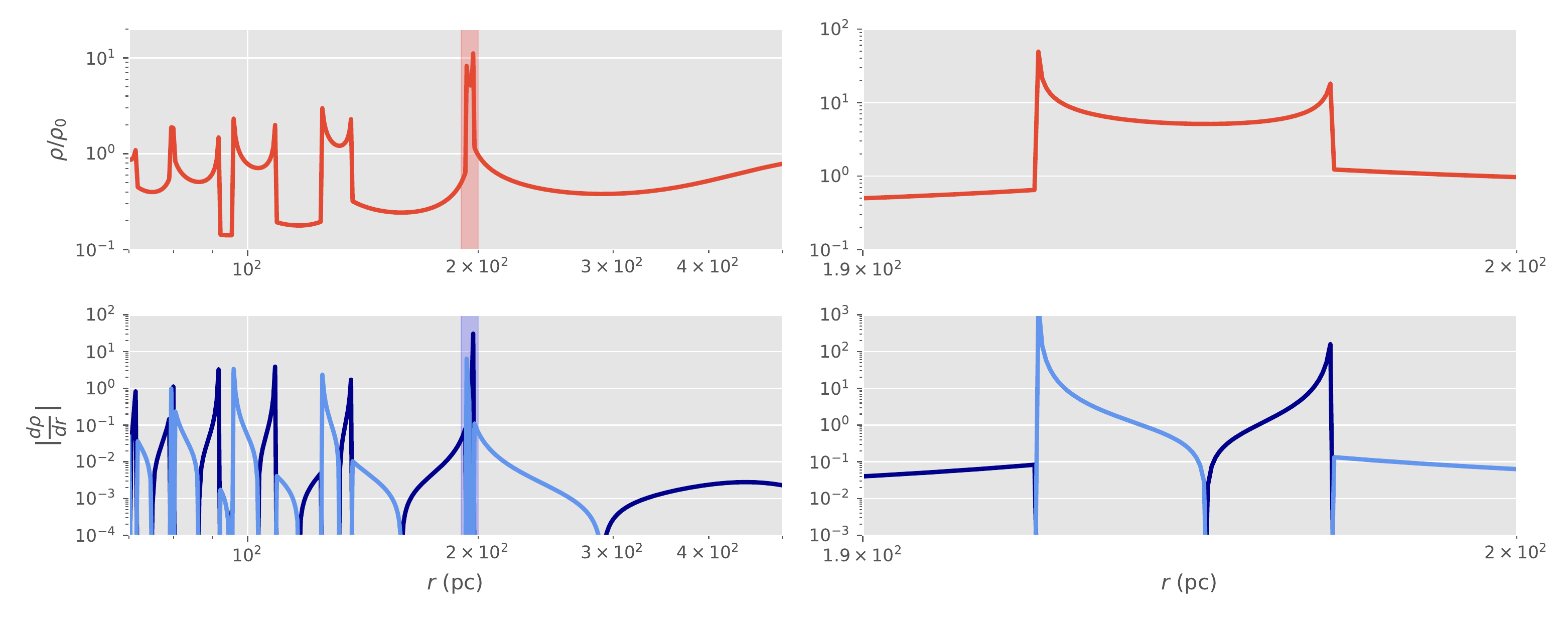}
\caption{The density profile (top, red) from equation \ref{rho} and
  its gradient (bottom, blue) from equation~\ref{drho_dr}. The left
  panels show a broad view of the profile, and the right panels zoom
  in on the outermost caustics (shown as shaded regions on the left).
  Positive gradients are shown in dark blue, and negative in light
  blue. The same parameters are used as in Figure~\ref{r_r0}.}
\label{rho_grad_r}
\end{figure*}

\begin{figure}
\includegraphics[width=\columnwidth]{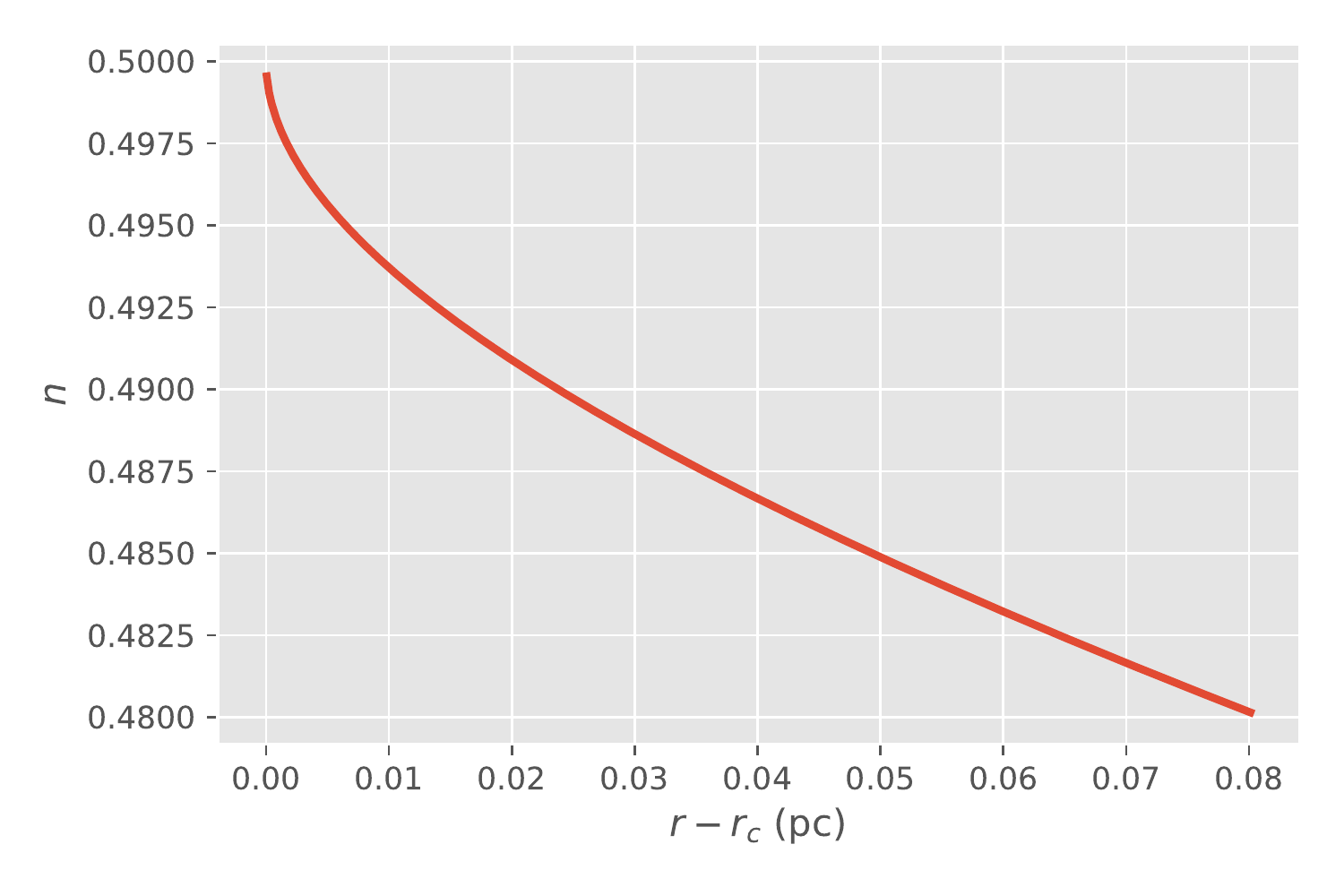}
\caption{The logaritmic slope $n$ of the density profile near a
  density peak at radius $r_c$ (defined in eq.~\ref{n}). As we
  approach the singularity, the slope tends to the limiting value
  $n=1/2$. The same parameters are used as in Figure~\ref{r_r0}.}
\label{n_r}
\end{figure}

In the caustics shown in Figure \ref{r0_rho_r}, the density is
formally singular, which suggests a potentially large (if not
infinite) interaction rate ($\propto \rho^2$) in such a system of
particles.  It can be shown, by expanding the expression for the
density close to the caustics, that as $\vert
r-r_c\vert/r_c\rightarrow0$, the density approaches the singularity as
\begin{equation}
\label{rho_close}
\rho(r) \propto (r-r_c) ^{-\frac{1}{2}}
\end{equation}
where $r_c$ is the location of the caustic. This derivation is shown
in Appendix \ref{Perturbation}.  To understand the profile from a
finite distance from the caustic, we can directly compute the gradient
of the density from equation \ref{rho},
\begin{equation}
\label{drho_dr}
\frac{d\rho(r)}{dr}=\sum_{r_i(r_0,t)=r} \frac{d\rho_i}{dr},
\end{equation}
where
\begin{equation}
\frac{d\rho_i}{dr} = \frac{1}{\mu^3} \left[\frac{dr_0}{dr}\frac{d\rho_0}{dr_0} + \mu^3 \rho_0(r_0) \frac{d\eta}{dr} \frac{d}{d\eta} \left( \frac{r_0^2}{r^2} \left\vert \frac{dr}{dr_0} \right\vert \right) \right]_{\eta_i}
\end{equation}
is the gradient in density for a single shell, with corresponding
$\eta_i$, at radius $r$. With tedious differentiation which we will
not reproduce here, this expression can be evaluated as a function of
$\eta_i$.  For completeness, we have included a pre-mass-loss gradient
$\frac{d\rho_0}{dr_0}$ here, although we for simplicity, we use a flat
profile ($\frac{d\rho_0}{dr_0}$=0) in our calculations. This is
justified by the fact that, near the caustic, the right hand term is
$O\left( \left\vert \frac{dr}{dr_0} \right\vert^2 \right)$ and
dominates over the $\frac{d\rho_0}{dr}\approx O\left(\left\vert
\frac{dr}{dr_0} \right\vert \right)$ term.

Figure~\ref{rho_grad_r} shows the profile and its gradient, and
includes a zoomed-in view near the outermost caustics. Notice that as
we zoom in and the numerical resolution increases, so does the height
of the caustics -- only numerical resolution keeps them from being
truly singular. We also note that the outermost turning point in
$\frac{dr}{dr_0}$ in Figure~\ref{r_r0} corresponds to the
second-largest-radius density peak in Figure~\ref{rho_grad_r}. (The
order of the caustics differs between
Figures~\ref{r_r0}~and~\ref{rho_grad_r}: they appear in pairs with the
larger $r$ corresponding to the smaller $r_0$.)

Let us assume a power-law density profile approaching a caustic from above,
\begin{equation}
\label{rho_c}
\rho_c \propto (r-r_c)^{-n},
\end{equation}
where $r_c$ is the radius of the caustic and $n$ is some power greater
than 0. Note that the sign of the term in brackets should be reversed
for peaks that approach the singular point from below (which is true
of every other peak).  In either case,
we can differentiate equation \ref{rho_c} to give
\begin{equation}
\label{n}
n\equiv -\frac{d\ln{\rho}}{d\ln{\vert r-r_c \vert}}= \frac{\vert r_c-r\vert}{\rho}\frac{d\rho}{dr}.
\end{equation}
and comparing this to the calculated density gradient we can find the
best-fit value of $n$ as the profile approaches the peak. This expression is
true for caustics which approach the singularity either from above or below.

Figure \ref{n_r} shows the value of this exponent near the
peak. Notice that it tends to $n=\frac{1}{2}$ as it reaches the
caustic (as expected from equation \ref{rho_close}). Immediately
away from the peak the profile becomes shallower.

\subsection{The particle interaction rate}
\label{caustic_rate}

\begin{figure}
\includegraphics[width=\columnwidth]{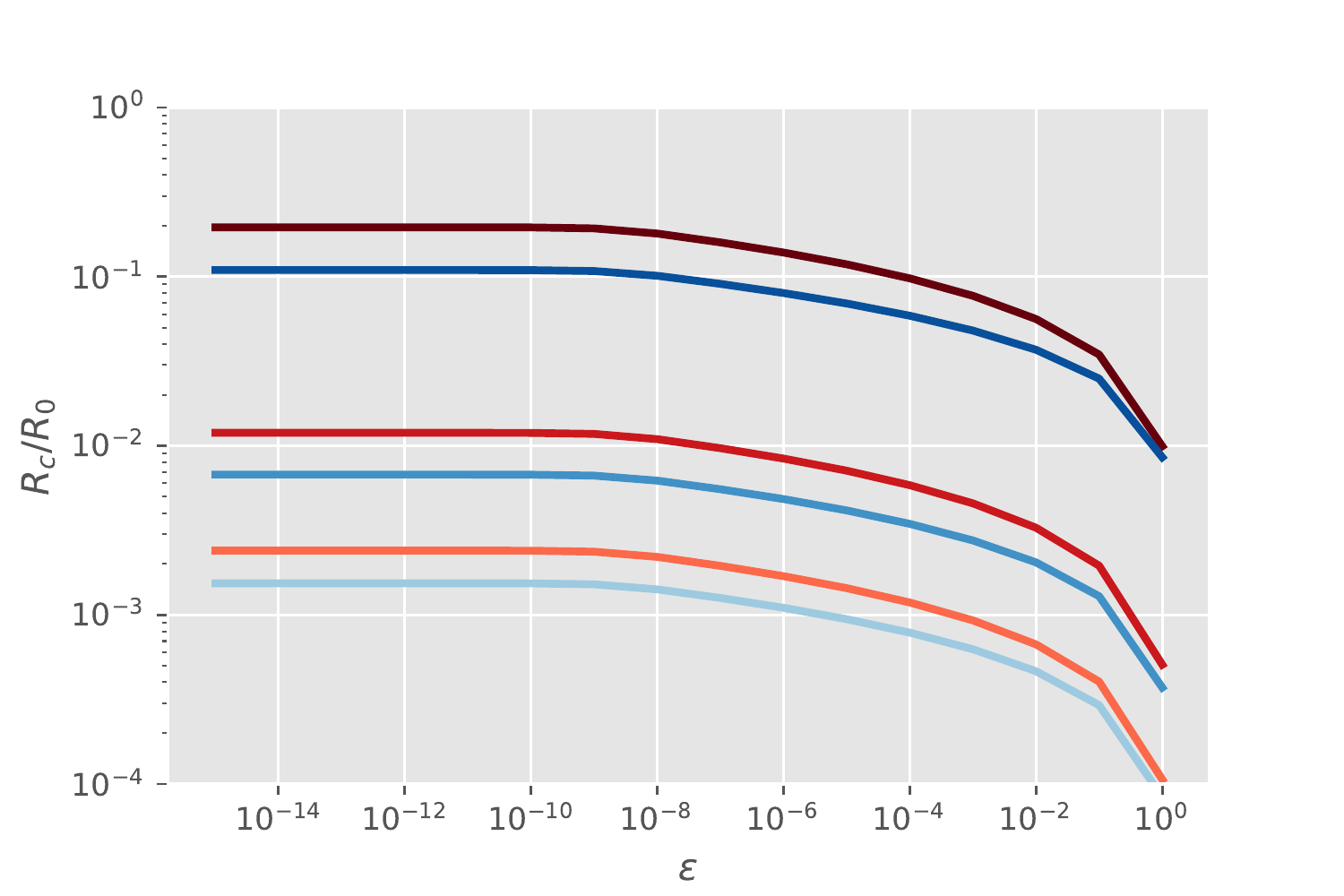}
\caption{The contribution to the interaction rate due to the caustics
  $R_c$ (equation \ref{caustic_boost}), compared to the rate over the
  whole system before mass loss $R_0$ (integrated to 500 pc, roughly
  the radius at which the perturbed density profile coincides with the
  unperturbed). The upper limit of integration, $\Delta$, is fixed at
  5pc (and the qualitative result is independent of this value) and
  the lower limit is varied to demonstrate numerical convergence. We
  perform the integral on the analytic density profiles of the first
  six caustics (counting in descending $r_0$). Red curves show the
  result for caustics which approach the singularity from above, and
  blue for those which approach from below. Darker curves show the
  contribution from caustics at larger radii. The integral is
  performed using Gaussian quadrature and the same parameters are used
  as in Figure~\ref{r_r0}.}
\label{boost_resolution}
\end{figure}

We now turn to the issue of whether this density profile, with sharp
(and formally singular) post-mass-loss density spikes leads to a large
boost in the particle interaction rate.

Assuming, for simplicity, a constant velocity
dispersion and interaction cross section, the interaction rate is
proportional to the integral
\begin{equation}
\label{rate_plain}
R \propto \int \rho^2 r^2 dr
\end{equation}
(these assumptions are discussed further in \S~\ref{Boost} below). 
We can calculate the contribution, $R_c$, to the total interaction
rate from a thin radial shell stretching from some small distance from
the caustic, $\varepsilon$, to some macroscopic distance, $\Delta$,
\begin{equation}
\label{caustic_integral}
R_c = \int^{r_c+\Delta}_{r_c+\varepsilon} \rho^2 r^2 dr.
\end{equation}
If the interaction rate over the caustic is finite then the value of
the integral should converge as $\varepsilon \rightarrow 0$.

Using the power-law form of the density near a caustic from equation
\ref{rho_c} with $x=r-r_c$ (and swapping all appropriate signs for
caustics which approach the singularity from below),
\begin{equation}
\label{caustic_boost}
R_c=\int^{\Delta}_{\varepsilon} k^2 x^{-2n} (r_c+x)^2 dx.
\end{equation}
For small $x$, the integrand approaches $k^2 r_c^2 x^{-2n}$, so that
the integral diverges for $n \ge \frac{1}{2}$ and $\varepsilon
\rightarrow 0$, implying an infinite net interaction rate. For $n <
1/2$, the integral is finite, though still potentially large, and
evaluates to
\begin{eqnarray}
\label{caustic_boost-solution}
R_c&=& k^2 x^{-2n}\left(\frac{x^3}{3-2n}+\frac{r_cx^2}{1-n}+\frac{r_c^2x}{1-2n}\right) \\
&=& \frac{k^2r_c^{3-2n}}{1-2n}\left(\frac{x}{r_c}\right)^{1-2n} \left[1+O\left(\frac{x}{r_c}\right)\right].
\end{eqnarray}
For $n$ close to $\frac{1}{2}$ the factor $\frac{1}{1-2n}$ can become
very large while the power of $x$ goes to 0 (and thus even for small
$x$ the integral can be large).

As the value of $n$ varies with radii this integral must be performed
numerically, and this result is shown in Figure
\ref{boost_resolution}. Here we see that the integral indeed
converges, and that the peaks at largest radii contribute the most to
the total interaction rate.

The total contribution to the interaction rate from the caustics,
found by summing over the first 100 peaks, is $R_{caustic}/R_0 =
0.23$, where the contribution from the inner caustics quickly becomes
vanishingly small. Thus the interaction rate of particles in caustics,
while very large for the small area the reside in, does not lead to a
net increase in the total interaction rate.

We can also integrate the rest of the profile separately, which is now
numerically feasible without having to resolve the caustics. The sum
of these two is a good approximation to the total ratio of $R$ to
$R_0$. Integrating the analytic profile, while capping the value of
$\rho$ at 100 (and thus not including the contribution from the
caustics) we find the contribution from the rest of the profile
$R_{profile}/R_0 \approx 0.25$.

Summing the interaction rates for the caustics and the rest of the
profile we find $R/R_0 \approx 0.5$. Hence the total interaction rate
following rapid mass loss is significantly less than the interaction
rate before mass loss.  In fact, as we argue in more detail in
Section~\ref{Other}, we have calculated this rate in an extremely
idealised case. Introducing more realistic physical effects will cause
these peaks to flatten, further decreasing the interaction rate.

It should be noted however that if the integration range does not
enclose the whole profile (specifically if it includes a pair of
caustics but not the associated density deficit at higher radii) then
the ratio $R/R_0$ can be larger than one. The only justification for
limiting the integration range as such would be if this was the outer
extent of the system, e.g. if a uniform density disk had a sharp
cutoff at a finite radius. Similarly if the integration range is taken
to be very large the ratio will tend to 1, regardless of the
mass-loss, as it will be dominated by mass at large radii which has
barely deviated from its original orbit due to its long period.

The integration limits used here are chosen to capture the full region
(minus the minor contribution at small radii, which we are limited
from resolving numerically) in which the density deviates from its
initial state.

Thus we have shown that despite the presence of a formal singularity,
the caustics themselves will provide only a minor contribution only to
the total interaction rate.  (A similar argument holds for disks and
other axisymmetric systems.) In fact the total interaction rate is
decreased by rapid mass loss, in direct contribution to the
expectation that these sharp density cusps may be an excellent
laboratory for observing high interaction rates.

As the shape of the profile is time-independent this result will hold
at all times, as long as the integration range is changed to encompass
the whole profile.

As an aside, we note that for three-body processes, whose rate is
$\propto\rho^3$, the rate near the caustics will diverge, and further
study into cases where these are physically relevant may be fruitful.

\subsection{Summary}

In this section, we have presented an analytic derivation of the
motions of shells of particles, initially on circular orbits,
following an instantaneous mass loss. We then found the corresponding
density profile numerically, and found the following properties:

\begin{itemize}
    \item The profile can be broadly split into two components:
    \begin{enumerate}
        \item Step-like over- and underdensities corresponding to
          regions where multiple shells overlap at one radius
        \item Singular caustics at radii where the edges of a single
          shell cross and hence its volume goes to 0 and its density
          to $\infty$
    \end{enumerate}
  \item As $r \rightarrow 0$, there is an infinite sequence of
    caustics, coming in pairs and spaced closer together at the edges
    of the regions where multiple shells overlap
  \item At large radii, particles have long periods, and well beyond
    the radius where the orbital time is longer than the time elapsed
    since the mass loss, the density profile tends to the unperturbed
    profile
  \item For a given fractional mass loss, the density profile is
    self-similar, with a shape that expands to larger radii
    as $r \propto t^{2/3}$ 

  \item The interaction rate in the caustics is large given their
    small spatial extent. Though the density profile is singular the
    caustics are shallower than the curve $\rho \propto
    r^{-\frac{1}{2}}$ and hence the integral of $\rho^2$ over some
    small region is finite. However the total interaction rate in the
    caustics is still small compared to the interaction rate of the
    profile preceding mass loss.

  \item Integrating over the whole profile, the interaction rate is
    less than the unperturbed case, a result that is independent of
    time.
\end{itemize}

\section{Response of less idealised systems}
\label{Other}

\begin{figure}
\includegraphics[width=\columnwidth]{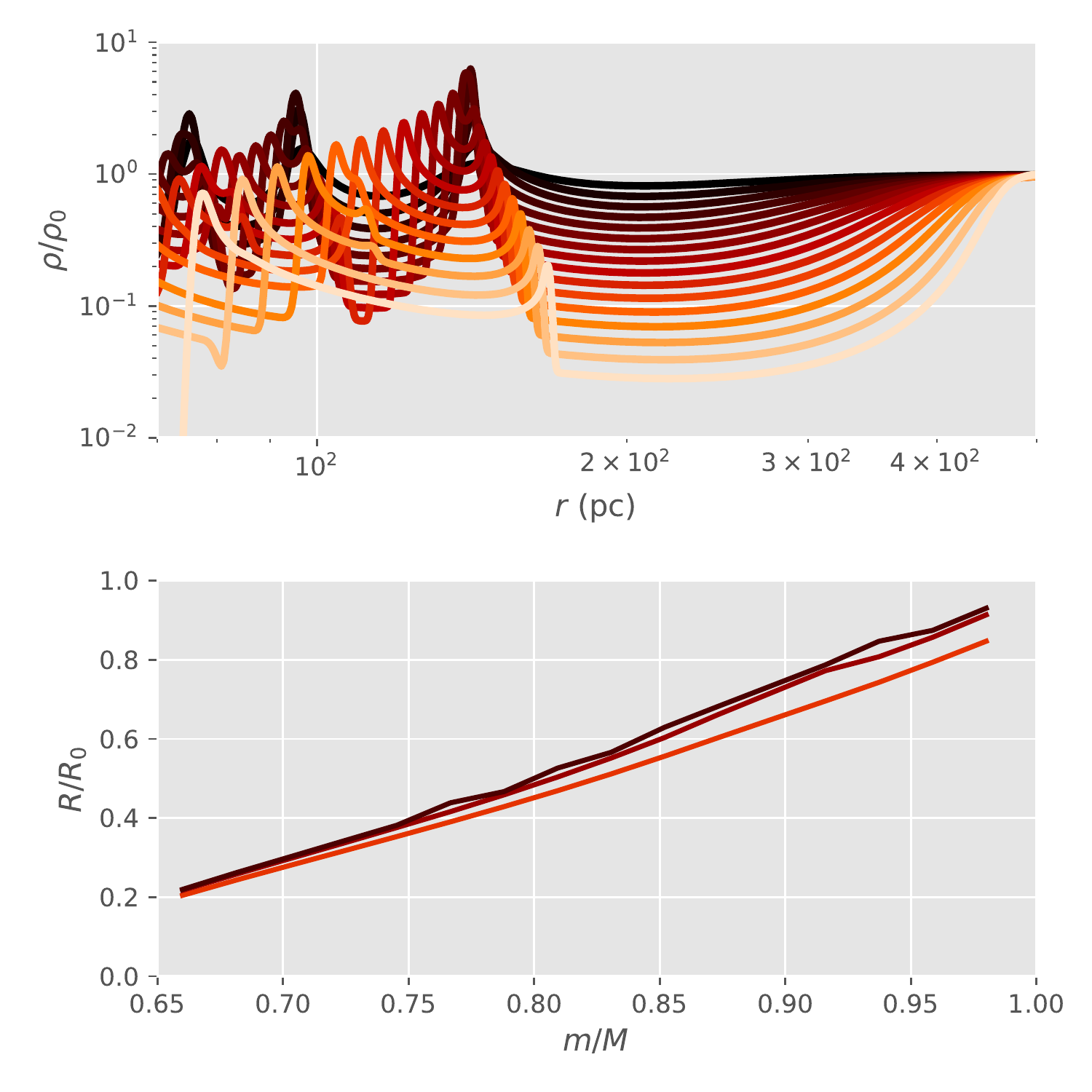}
\caption{{\it Upper panel:} The density profile 6 Myrs after
  instantaneous mass loss, shown for a range of values of $m<M$ for a
  constant $M=10^9 M_\odot$. From dark to light, $m/M$ goes from 0.98
  to 0.66 in intervals of 0.02. The profile has been computed using
  the numerical CausticFrog package and sharp peaks of the caustics
  have been smoothed over 1 pc (more details in Section~\ref{Other}
  and Appendix~\ref{Code}). {\it Lower panel:} The total interaction
  rate, $R$, compared to the rate before mass loss, $R_0$ (calculated
  by numerically integrating the above smoothed profiles from 70 to
  500 pc using equation \ref{rate_plain}). Three different smoothing
  lengths are used: 5 pc (light red), 1 pc (dark red) and $10^{-10}$ pc
  (black), to show that the results are very weakly dependent on
  smoothing length.}
\label{rho_dm}
\end{figure}

In \S~\ref{Kepler}, we chose the simplest physical system, consisting
of massless particles initially on circular orbits in a Keplerian
potential, so as to find a semi-analytic density profile. Here we
discuss results from a range of more physical realisations. We show
that each amendment leads to a flatter, less sharply contoured,
profile than the circular orbit case.

Whenever density profiles are shown, they have been found with our new
public 1D Lagrangian simulation code \textsc{CausticFrog}. By evolving
the edges of a series of spherical shells, which are able to cross and
overlap, we can easily resolve both shell crossing and squeezing (and
hence resolve caustics) exploiting the spherical symmetries of the
problem to save computation costs. This code can be found at
\url{https://github.com/zpenoyre/CausticFrog} and is presented in
detail in Appendix \ref{Code}.

Numerical discreteness noise, caused by the thousands of interacting
shells, makes these profiles difficult to inspect visually and to
integrate over numerically, so we smooth the profile.  This is done
by replacing each spherical shell, extending from radius $r_1$ to
$r_2$ with a Gaussian centred centered at $\frac{1}{2}(r_1 + r_2)$ and
with a width
\begin{equation}
    \sigma = \sqrt{\frac{1}{4}(r_2 - r_1)^2 + r_s^2},
\end{equation}
where $r_s$ is a smoothing length. The profile is normalised to
conserve mass.  Note that the realistic density profiles we consider
below consist of convolving a set of discrete caustics with a smooth
distribution.  We thus expect the caustics structures to be physically
smoothed, justifying this approach.  Furtheremore, in \S~\ref{rho_dm}
we show that the choice of smoothing length does not have a large
impact on the interaction rate, and hence does not qualitatively affect
the results.

\subsection{Degree of mass loss}
\label{total_rate}

We have shown in Section \ref{Kepler} that the density profile
resulting from a specific drop in a point mass potential ($10\%$ for
all above analysis) leads to a drop in the total interaction rate of
particles in the system. Now we extend this to any fractional mass
drop.

Figure \ref{rho_dm} shows the response of the Keplerian system with
initially circular orbits to varying degrees of mass loss. The top
panel shows the density profiles at the fixed time 6 Myr following a
mass-loss of various degrees.  Clearly, more significant mass loss
leads to lower overall densities, as the particles are significantly
less tightly bound to the smaller remaining point mass. Thus they have
more eccentric orbits and move further outwards, giving a lower
density. Smaller mass losses lead to flatter and less perturbed
profiles.

The bottom panel in Figure \ref{rho_dm} shows the ratio of the
particle interaction rates 6 Myrs after ($R$) {\it vs} immediately
before ($R_0$) the moment of mass loss.  The interaction rate is
reduced, regardless of the degree of mass loss. The larger the mass
loss, and thus the more eccentric and larger the orbits of the
particles, the lower the density and the lower the interaction
rate. Note that the particles become unbound for mass losses of
$50\%$, though the interaction rate will be non-zero for some period
while the unbound mass moves outwards.

Hence the drop in interaction rate is true for any fractional mass loss.

\subsection{Self-gravity}
\label{self_gravity}

\begin{figure}
\includegraphics[width=\columnwidth]{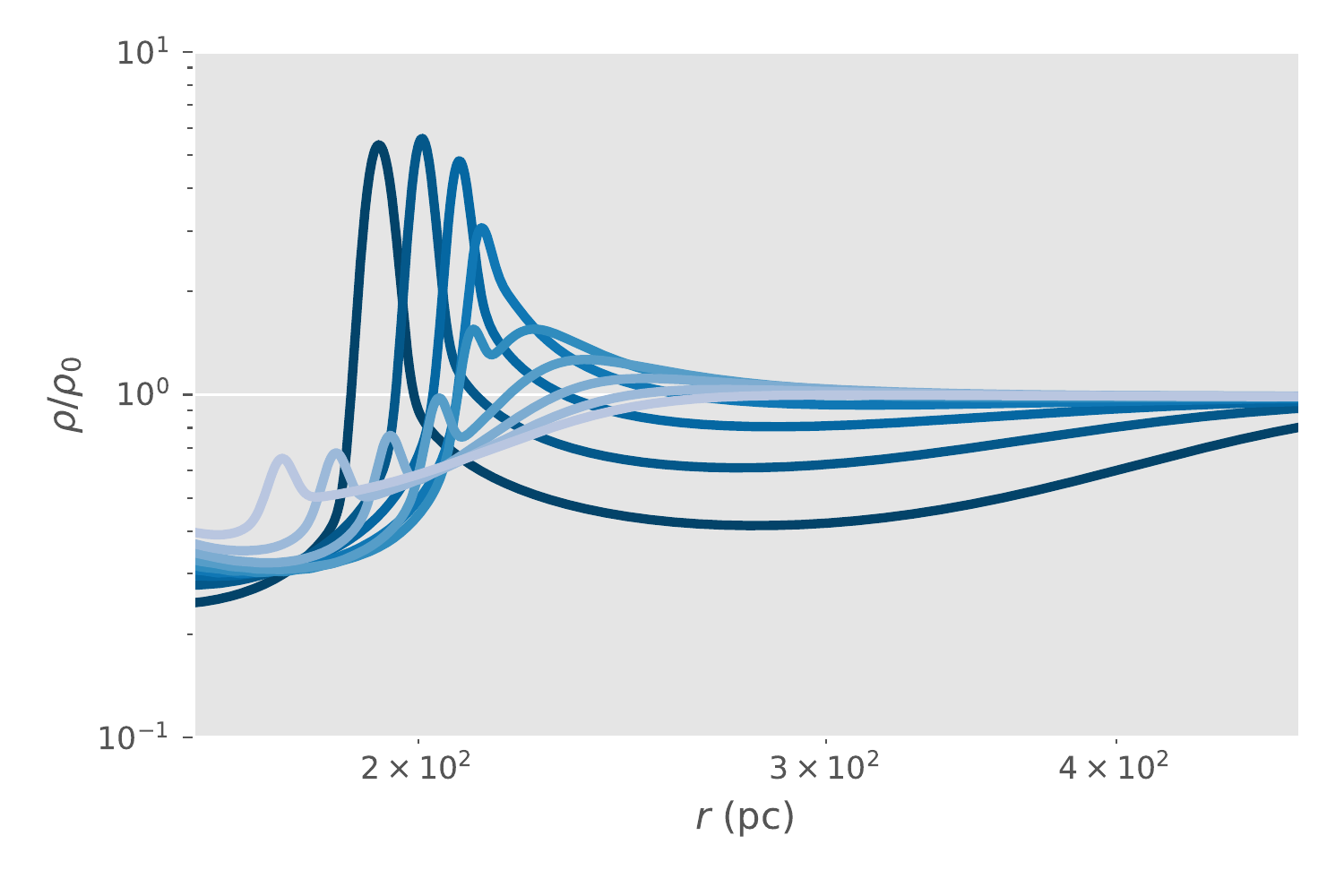}
\caption{The density profile around the outermost peak, at the same
  moment in time, for a range of initial densities (and hence
  contribution via self-gravity). All profiles are smoothed over a 2
  pc scale. From dark to light the densities increase from 1 (visually
  indistinguishable from the analytic solution) to 33
  $\mathrm{M}_\odot \mathrm{pc}^{-3}$ (in steps of 4 $\mathrm{M}_\odot
  \mathrm{pc}^{-3}$). The same parameters are used as in Fig.~\ref{r_r0}.}
\label{rho_self}
\end{figure}

Depending on the system in question, self-gravity may be safe to
ignore (e.g. the inner regions of accretion disks around black holes),
or it may be the dominant source of gravitational potential (e.g. the
dark matter profile away from the centre of a gas-poor dwarf galaxy).

Equation \ref{scale} shows that the speed at which the contours of the
density profile move outwards depends on the enclosed mass. In a
self-gravitating system, as a feature moves outwards, the mass enclosed
generally increases; hence, the speed increases and the profile spreads
out. Equation \ref{scale} is no longer exact with the inclusion of
self-gravity (and hence unclosed orbits), but in the case of a central
point mass it is an increasingly good approximation as the initial
density goes to 0.

We next explore the impact of self-gravity with \textsc{CausticFrog}
by following a system of particles on initially circular orbits, as
before, but including the self-gravity of each shell. We examine the
profile for the mass losses and time periods as in the analytic case
(see Fig.~\ref{r_r0} for details).

The results of this exercise are shown in Figure~\ref{rho_self}, for
initial profiles with increasing density (and hence contribution of
self-gravity).  The figure shows that denser systems have features
that move outward faster, reaching larger radii at a given time, with
more dispersed peaks.  When the mass of gravitating fluid enclosed,
$m_{enc}$, is of order of the mass of the central object (as is true
for the lightest curve) the caustic is almost entirely dispersed. Even
when $m_{enc} \sim 0.1 m$ (darker curves) the difference between the
point mass and the self-gravitating profiles starts to become
apparent.

We conclude that self-gravity will generally disperse caustics and
lead to smoother and flatter density profiles.

\subsection{Non-circular orbits}
\label{non-circular}

\begin{figure}
\includegraphics[width=\columnwidth]{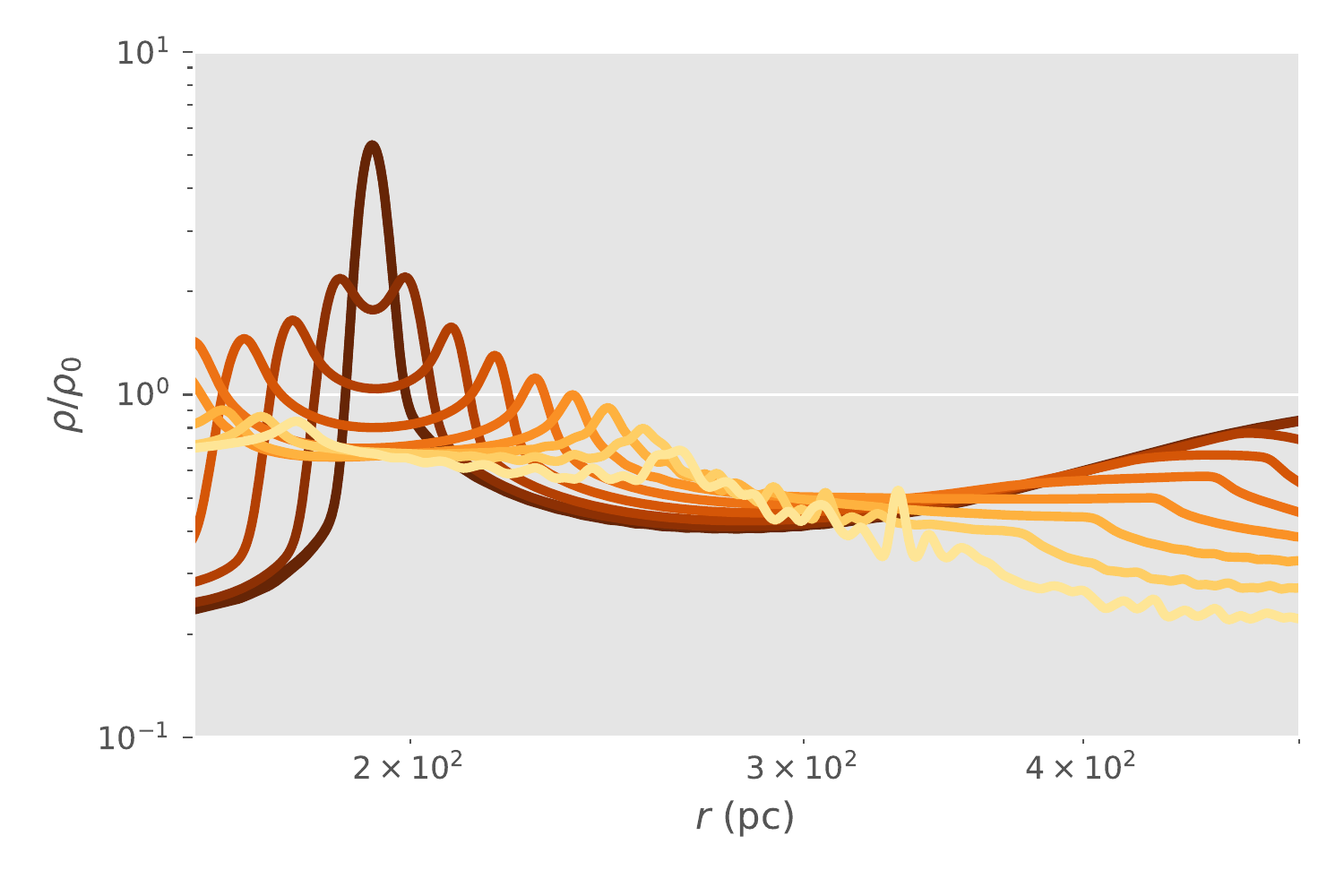}
\caption{The density profile near the outermost peak, at the same
  moment in time, for different initial eccentricities. From dark to
  light the eccentricities range from $e=$ 0 to 0.4 (in steps of
  0.05). The same parameters are used as in Fig.~\ref{r_r0}, and the
  curves are smoothed over 2 pc (for the highest eccentricities there
  are few enough remaining bound shells that even this does not smooth
  out all numerical noise).}
\label{rho_ellipse}
\end{figure}

\begin{figure}
\includegraphics[width=\columnwidth]{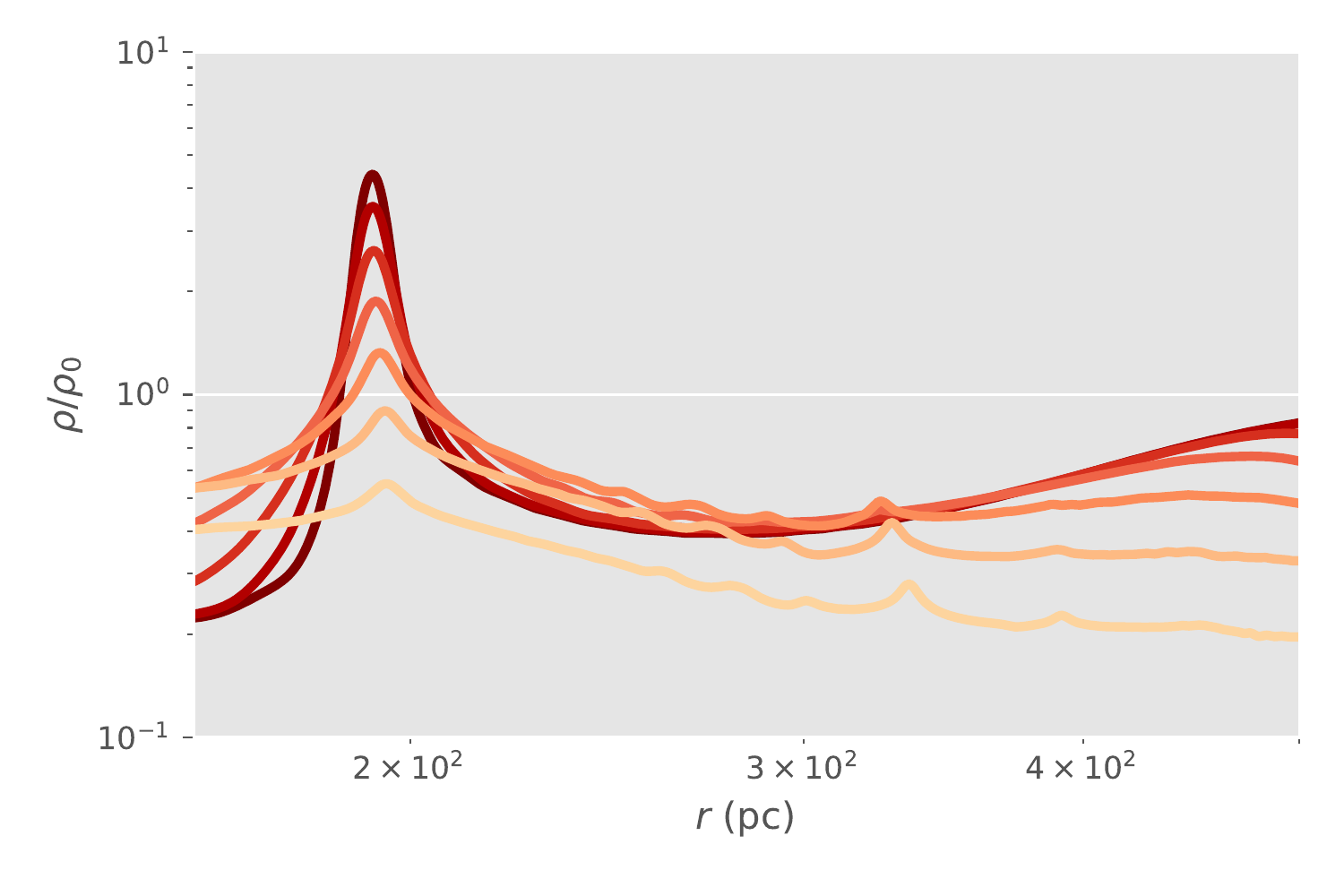}
\caption{The post-mass loss density profile for a range of initial
  eccentricities. The initial eccentricity distribution is given by
  Eq.~\ref{p_ecc} with different values of $n$ controlling its width,
  from nearly circular (large $n$) to broader distributions including
  high eccentricity (small $n$).  From dark to light, the curves
  correspond to $n=$ 64, 32, 16, 8, 4, 2, and 1 respectively. The same
  parameters are used as in Fig.~\ref{r_r0}, and the curve is smoothed
  over 2 pc.}
\label{rho_many}
\end{figure}

For most systems, we would not expect the gravitating particles to be
on circular orbits. In some cases, such as in a gas disk, viscous
dissipation may circularise orbits, but in a dissipationless system
such as a dark matter halo or a stellar bulge, we expect a wide
distribution of orbital eccentricities.

In this section, we present numerical solutions for non-zero initial
eccentricities using \textsc{CausticFrog}, but many of the results
below are equally apparent from the analytic derivation presented in
Appendix \ref{EllipseKepler}. In particular, particles with different
phases at the moment of mass loss will reach their turning points at
different times. We expect that this should cause sharp features of
the density profile near caustics to spread out.

Let us start with the simple case of orbits of a fixed initial
eccentricity. We would expect the distribution of initial phases to
correspond to $p(\phi) \propto \dot{\phi}^{-1}$ (the probability of a
particle being at some phase is inversely proportional to the rate of
change of phase), and initialise the initial positions of particles
along their elliptical orbits accordingly.  We simulate just over a
million such orbits.

Figure \ref{rho_ellipse} shows the density profile for orbits with
different initial eccentricities (and a full range of initial phases),
between $0\leq e\leq 0.4$. As we move to higher and higher
eccentricity, the density peaks split into multiple peaks, and the
density profile flattens overall. At high eccentricities, a
significant amount of mass is unbound after mass loss and the density
drops precipitously.  We next compute the density profiles in a more
realistic situation, for initial orbits with a wide range of initial
eccentricities and phases. We use a simple toy model of the
distribution of initial eccentricities,
\begin{equation}
\label{p_ecc}
p(e) \propto (1-e)^n,
\end{equation}
where $n$ can be chosen to give mostly circular orbits (large $n$) or
a much wider range of eccentricities (small $n$).  We emphasize that
this distribution is ad-hoc, but it conveniently allows us to explore
the impact of the width of the initial eccentricity-distribution.
Figure \ref{rho_many} shows density profiles for a range of values of
$n$. When the distribution is sharply peaked around $e=0$, the
resulting profile still has sharp spikes, but for broader
distributions, those peaks are much smaller. When a large fraction of
high-eccentricity orbits are included, a significant amount of mass
can again be lost, as particles become unbound.

We conclude that a system with a wider range of initial eccentricities
will have smoother features, and a flatter overall density profile,
following a mass loss event. For sufficiently large mass loss, there
is a net reduction in mass as particles initially near their periapsis
can easily become unbound (see Eq.~\ref{ecc_unbound}).

\subsection{Other assumptions and approximations}

There are several additional complications that could change the
response of a system to rapid mass loss. Here we briefly discuss a
few of these complications qualitatively.

\subsubsection{Time dependence}

The basic premise of this system is that mass loss is almost
instantaneous, i.e. occurs on a timescale that is much shorter than
the particles' orbital time.  While instantly removing the mass makes
our calculation much easier, allowing for the mass to decrease over a
finite (if short) period will smooth out peaks and further flatten the
density profile.

A simple way to picture this is to imagine the mass dropping not in a
single event, but two curtly spaced events. In an initially circular
case, the first event sets particles onto elliptical orbits. When the
second event occurs, particles are on a range of orbits with different
phases. We can think of
this a second 'initial' state, now with particles with different
orbital properties at the same radius. As shown in
\S~\ref{non-circular}, a system with a variety of initial orbits
generally has less strongly peaked features than a family of similar
initial orbits. Hence, the profile will be flatter than if the mass
had dropped in a single event. This argument could be extended to
reducing a single mass loss to any number of distinct steps, and hence
to a continuous mass loss rate.

The results shown above for a self gravitating fluid (\S~\ref{self_gravity}) can also be understood as a time dependant phase mixing, as now the orbit of one particle (or spherical shell) directly affects another and over time they exchange energy. Thus the caustic, a region of high or even infinite phase density, diffuses and flattens over time.

\subsubsection{Alternative potentials}

We expect most physical potentials to have some degree of asphericity
\citep{Pontzen15} which will break the spherical symmetry of our
solutions. Relativistic effects may also be important;
relativistic precession, for example, will also disrupt any simple dependence
between an orbit and its period. Furthermore, as shown in
\S~\ref{self_gravity}, the inclusion of self-gravity will also break
down sharp features, and thus any self-consistent profile (such
as the profile for dark matter halos suggested in \citep{Navarro97})
cannot maintain strong features.

\subsubsection{Dissipation}

We have so far assumed a collisionless system -- but, depending on the
context, there are several ways for the post-mass loss density waves
to dissipate energy. We expect that such dissipation will spread the
initially highly coherent waves, and the profile will flatten as a
result.

For baryonic matter, viscous dissipation due to turbulence and
magnetic fields, or due to radiative processes, can all
be effective at sapping energy from dense, fast-moving
regions. Furthermore, shocks can develop as the overdensities move
outwards, heating and transferring energy to the medium they move
through.  Finally, pressure forces generally smooth the perturbations
caused by the mass loss \citep[see e.g. the discussion in][and
  references therein]{Corrales10}.

\subsubsection{Unbound mass}

We have already seen that in systems with highly eccentric initial
orbits, only small changes in the central mass are needed for some
particles to become unbound.  Any mass loss will of course lead to a
lower density, and this will further flatten the profile.

\subsection{Summary}

We have discussed some effects that should be incorporated
quantitatively into a more realistic picture of a dynamical system
before and after a period of rapid mass loss.  A general trend is
clear: compared to the simplest idealised case presented in
\S~\ref{Kepler}, a more physical model develops a smoother and flatter
density profile.  This will generally reduce the particle interaction
rate compared to the idealised case.
     
\section{Interaction Rates - A General Discussion}
\label{Boost}

In the one case for which we have calculated the interaction rate
(initially circular orbits in a Keplerian potential, \S~\ref{Kepler})
we have shown that there will be a smaller
interaction rate than before the mass loss event, as we have observed
in \S~\ref{total_rate}. Here present a more general
heuristic argument: namely, if mass on average moves outwards (as
is the case following mass loss), the interaction rate will generically
decrease.

First let us more carefully justify our calculation of the particle
interaction rate. The rate per unit volume, for a single fluid with a
Maxwell-Boltzmann velocity distribution, is $\propto n^2 \sigma
\sqrt{\langle v^2 \rangle}$ where $n$ is the number density, $\sigma$
the interaction cross section and $\langle v^2 \rangle$ the velocity
dispersion.

We will make the simplifying assumptions that (i) the cross section is
constant throughout and (ii) the velocity dispersion is unchanged
before and after mass loss.  We have so far not specified the
orientation of the initial orbits. Two limiting cases are isotropic
initial velocities for randomly inclined orbits, or zero dispersion if
all orbits are co-planar and in the same direction. In the isotropic
case, the assumption of constant velocity dispersion before and after
mass loss is reasonable, but for anisotropic initial velocity
structures, this assumption may break down.

We note that as mass loss induces particles to move outward on
average, their velocities are generally lower than the initial
velocities. We have now introduced a radial velocity dispersion as
shells moving radially cross, so the assumption that velocity
dispersion is unchanged (or reduced) is equivalent to the assumption
that the newly introduced radial dispersion is smaller than the
original tangential velocity dispersion.

To characterise the change in the total interaction rate of a system, we
define the ``boost factor" as the ratio of the interaction rate at a given
post-merger time to that before the moment of mass loss,
\begin{equation}
\label{boost}
B(t)=\frac{\int \rho_1^2(t) r^2 dr}{\int \rho_0^2 r^2 dr}
\end{equation}
where $\rho_0$ and $\rho_1(t)$ are the density profiles before and
after mass loss. For the total interaction rate of the system the
integral should be evaluated out to the radius of the system and can
be converted to a luminosity (for a given DM particle cross section)
to compare to observations.  We will assume here that we are
interested only in the integrated interaction rate, because the system
is unresolved. This is because we are dealing with small objects at
extragalactic distances (AGN disks) and/or because the actual signal
(e.g. gamma-rays from DM annihilations in dwarf galaxy cores) is
spatially unresolved.

We will also assume that mass is conserved as the density profile is
modified, i.e.
\begin{equation}
\label{mass}
M=\int 4 \pi \rho r^2 dr = \mathrm{const.}
\end{equation}
This integral will also extend to the outer edge of the system. As we
have seen in \S~\ref{Other}, mass can become unbound and lost, but
this will only reduce the densities and lead to smaller boost factors.

\subsection{General transformations}

\subsubsection{Change in volume}

\begin{figure}
\includegraphics[width=\columnwidth]{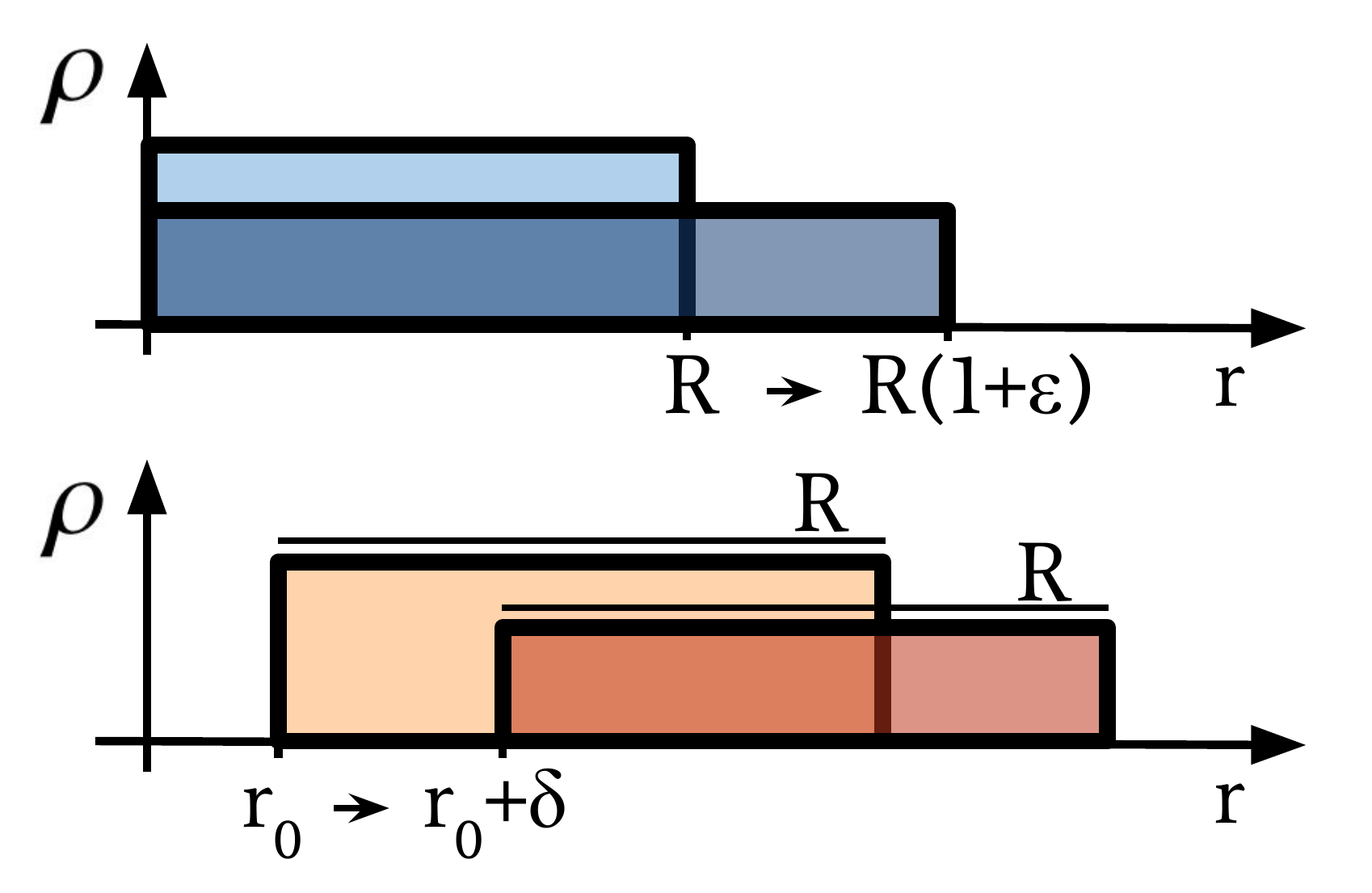}
\caption{A flat density profile is stretched (top panel), and shifted
  (bottom panel), both whilst conserving mass. This leads to a drop in
  the interaction rate.}
\label{combinedTransformation}
\end{figure}

Stretching the initial profile (see top panel of
Figure~\ref{combinedTransformation}) such that the outer radius is
some factor $(1 + \epsilon)$ times its initial value yields the new
density $\rho_1=(1+\epsilon)^{-3}\rho_0$, and thus the boost\footnote{In both this and the following calculation we have integrated over the whole profile. Integrating without changing the limits will lead to yet lower value of $B$.}
$B=(1+\epsilon)^{-3}$, i.e. $B<1$. The bottom panel of Figure~\ref{combinedTransformation} illustrates
another volume-expanding operation: shifting an initially flat density
profile to higher radii. The width of the profile, $R$, is unchanged,
and we use the dimensionless parameters $\alpha=\frac{R}{r_0}$ and
$\beta=\frac{\delta}{r_0}$ to describe the transformation. Conserving
mass leads to the density, and therefore the boost, dropping by a factor
\begin{equation}
\label{stretchRho}
\frac{\rho_1}{\rho_0}=B=\frac{(1+\alpha)^3 -1}{(1+\alpha+\beta)^3 - (1+\beta)^3}.
\end{equation}
For $\beta$ greater than 0 ($\alpha$ is always $> 0$), this again leads to $B<1$.

\subsubsection{Change in mass distribution}
\label{slope}


\begin{figure}
\includegraphics[width=\columnwidth]{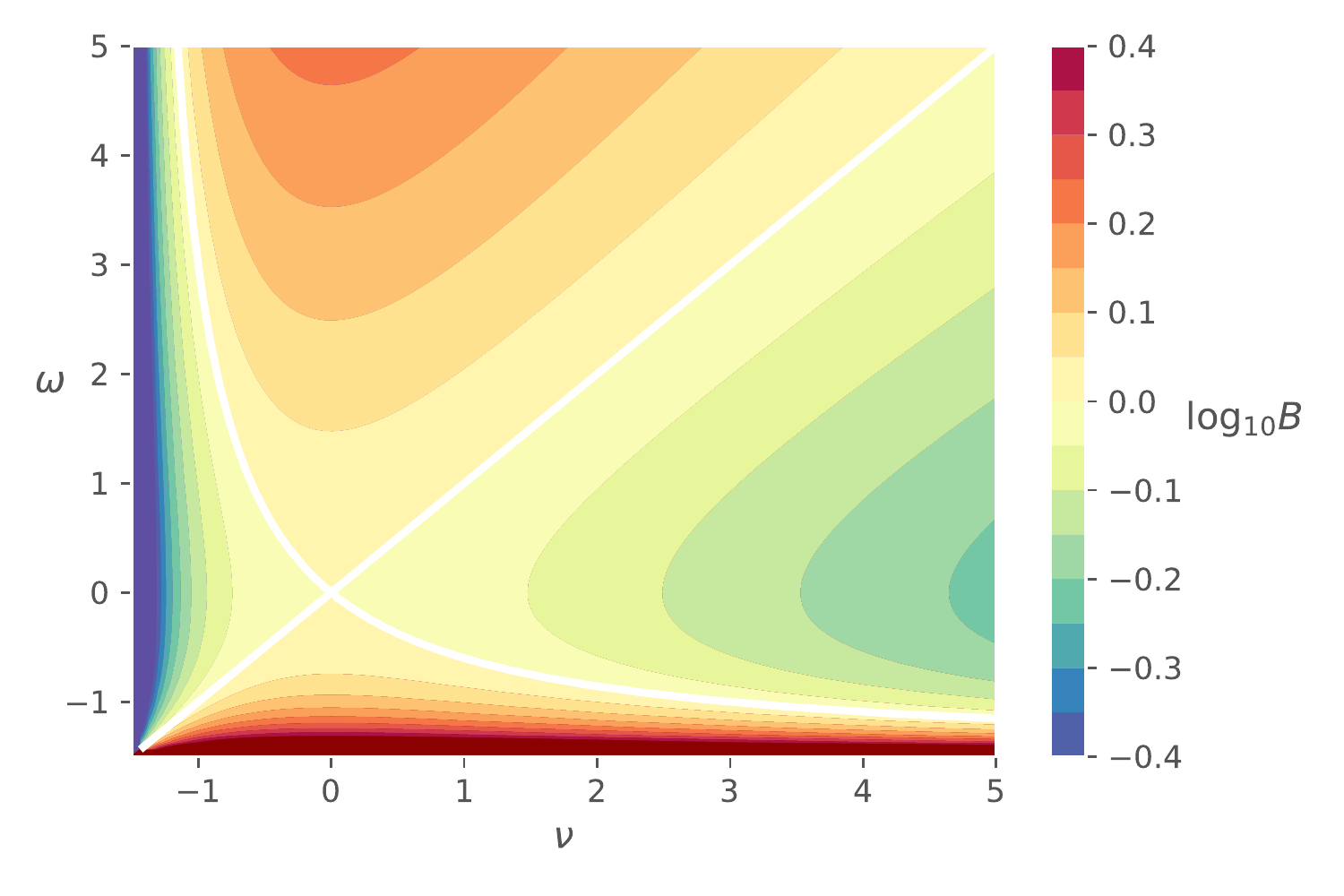}
\caption{The boost factor, as given by equation \ref{powerBoost}, if
  the density function goes from a power law with power $\nu$ to one
  with power $\omega$ (c.f. equation \ref{initPower}). The curves for which $B=1$ are shown in white. The
  profile extends to infinity for positive $\nu$ and $\omega$ but must
  be curtailed at $\nu=-\frac{3}{2}$ ($B \rightarrow 0$) and
  $\omega=-\frac{3}{2}$ ($B \rightarrow \infty$).}
\label{power_boost}
\end{figure}

We also examine the effect of a more general transformation to the
density profile: changing from one power law to another. Even if the
profile has many features, if smoothed, or averaged over time, the
profile will be well described by a simple power law.
Let us assume an initial and final profiles of the form
\begin{equation}
\label{initPower}
\rho_0 = k r^\nu \ \mathrm{and} \ \rho_1 = \kappa r^\omega,
\end{equation}
where both extend to the same outer radius, $R$. Conserving mass gives a boost
\begin{equation}
\label{powerBoost}
B = \frac{3+2\nu}{3+2\omega} \left( \frac{3+\omega}{3+\nu} \right)^2.
\end{equation}
Figure \ref{power_boost} shows how the boost varies with the initial
and final power law. If both powers are of the same sign, the boost is
less than unity if $\vert \omega \vert < \vert \nu \vert$, i.e. if the
resulting power law is shallower than the original, as we might expect
for mass becoming less bound and moving outwards.

When the power law changes sign the behaviour is more complex, with
the boost going to infinity as the power approaches $-\frac{3}{2}$. If
either power is negative it tends to dominate, unless the other is
very large and positive.

An important feature to note is that the largest boosts are seen along
the line $\nu=0$, an argument that the case presented in
\S~\ref{Kepler}, with an initially flat density profile, produces the
largest boost (though more complex families of solutions with larger boosts may still exist).

Thus if mass moves outward and a density profile flattens, the boost
decreases.

\subsection{Combining a smooth profile with caustics}

As we have seen in Figure \ref{r_r0}, even when mass in
general moves outwards, there can be small regions where the opposite
happens: the mass is squeezed into a smaller volume, or the density
profile steepens.

This of course is the cause of the caustics, as some finite mass is
squeezed into an infinitesimal volume. We have shown numerically, for
initially circular orbits around a point mass, in Section \ref{Kepler}
(and extended it to more general situations in Section \ref{Other})
that rapid mass loss does not lead to a boost in the interaction rate
in a system.

In other words, the global phenomenon, of mass becoming less bound and
moving outwards, dominates over the local phenomenon, of sharp density
peaks developing in small regions.

Moving away from a flat initial density profile, as shown in Section
\ref{slope}, will further decrease the boost. Furthermore, switching
to a self gravitating system, discussed in Section \ref{self_gravity},
flattens out the caustics and reduces the interaction rate. Thus we
can generalise this result for other astrophysical potential and density
profiles.

\section{Conclusions}
\label{sec:conclude}

First, let us briefly summarise the argument presented in this paper:

\begin{itemize}

\item A system which develops a large overdensity seems a strong
  candidate for observing large particle interaction rates ($\propto
  \rho^2$).

\item Rapid mass loss in a system leads to instantaneously changed
  orbits and the development of over- and underdensities as orbits
  cross and overlap.

\item After mass loss in a Keplerian potential, the density profile of
  particles on initially circular orbits is a combination of
  infinite-density caustics and step-like over- and underdensities.

\item The caustics in the circular Keplerian case contribute only a
  small amount to the interaction rate, significantly less than the
  total interaction rate before mass loss.

\item Away from the caustics, the step-like profile leads to a drop in
  the interaction rate as mass moves outwards (as particles are less
  bound after mass loss) and the density drops.

\item Overall, rapid mass loss in the circular Keplerian case leads to
  a \textit{smaller} interaction rate than the unperturbed case.

\item The inclusion of less idealized physical effects smooths and
  flattens the density profile relative to the circular-orbit
  Keplerian case. Mass still moves outward and thus the total
  interaction rate is reduced.

\item Hence rapid mass loss will, in \textit{any} physical case, lead
  to a drop in the interaction rate, rather than an increase.

\end{itemize}

Below, we elaborate upon how we arrived at this somewhat surprising
conclusion.

In \S~\ref{Kepler}, we present an analytic derivation of the response
of a system of particles, initially on circular orbits, in a Keplerian
potential to an instantaneous drop in the central point mass. From
this we numerically derive a density profile (Figures \ref{r0_rho_r}
and \ref{rho_grad_r}) that has a self-similar shape and expands
outward with time as $r \propto t^\frac{2}{3}$. The profile is
comprised of step-like over- and underdensities where multiple shells
on different orbits overlap, and singular caustics at the boundaries
of the multi-shell regions, where a finite mass is squeezed into an
infinitesimal volume.

These sharply peaked profiles with singular caustics naively appear
promising for a large increase in the total interaction rate.  However
we show that the rate in this case is \textit{still} less than in the
unperturbed case.  The caustics can be shown to contribute only a
small amount to the interaction rate, and regardless of degree of mass
loss or time (as the shape of the profile is time independent) the
interaction rate decreases.

In \S~\ref{Other}, we show that various effects to make the system
more realistic (such as self-gravity, non-circular initial orbits, and
non-Keplerian potentials) smooth out the sharp density spikes and lead
to flatter overall density profiles.  Thus, the circular Keplerian
case provides the profile that is most sharply peaked.

Thus we have shown that even the best possible candidate environment
for observing large interaction rates following rapid mass
\textit{still} has a smaller net interaction rate than the same system
before mass loss.

Similar to our results here, the optically thin Brehmsstrahlung
luminosity, computed in post-merger binary black hole accretion disk
simulations of \citet{O'Neill09}, \citet{Megevand09} and
\citet{Corrales10} have been found to {\it decrease} after the
mass-loss caused by the BH merger.\footnote{As explained in
  \citet{Corrales10}, this Brehmsstrahlung luminosity is not
  self-consistent, as it yields an unphysically short cooling time.
  Nevertheless, this luminosity involves an integral of $\rho^2$ over
  volume, and its post-merger decrease can be traced to the reasons we
  identified in this paper: the overall decrease in the emission due
  to the expansion of the disk dominates over the increased emission
  from the dense shocked rings.}
  
Our results -- the absence of a large boost in the
particle interaction rate -- also justify the simple density profiles
used to calculate $\gamma$-ray flux from dark matter annihilation in
dwarf galaxies \citep[e.g.][and references therein]{Geringer15}.

We emphasise that the arguments presented here are generalisable and
thus applicable to any other system or geometry where we observe
mass-loss over a period much shorter than the dynamical time of
orbiting particles. 

There are extreme cases where the interaction
rate may increase, such as if three-body interactions are the main
source of the signal, or where the step-like behaviour of the density
function is precipitously steep. The large densities in the caustics may also lead to other observable phenomena, such as due to the heating of gas in an AGN disk, but we leave these considerations for future work.

But the overall conclusion of this
work is that rapid mass loss in dynamical systems is \textit{not} the
promising laboratory for observing high interaction rates as one may
have hoped for.

\section*{Acknowledgements}

The authors thank Andrew Pontzen, Jacqueline van Gorkom, George Lake,
Nick Stone and the anonymous referee for their insightful comments and
queries, and Emily Sandford for invaluable help with the text.  This work
was supported in part by NASA grant NNX15AB19G; ZH also gratefully
acknowledges sabbatical support by a Simons Fellowship in Theoretical
Physics.




\bibliographystyle{mnras}
\providecommand{\noopsort}[1]{}



\appendix

\section{Non-circular orbits in a Keplerian potential}
\label{EllipseKepler}

\begin{figure}
\includegraphics[width=\columnwidth]{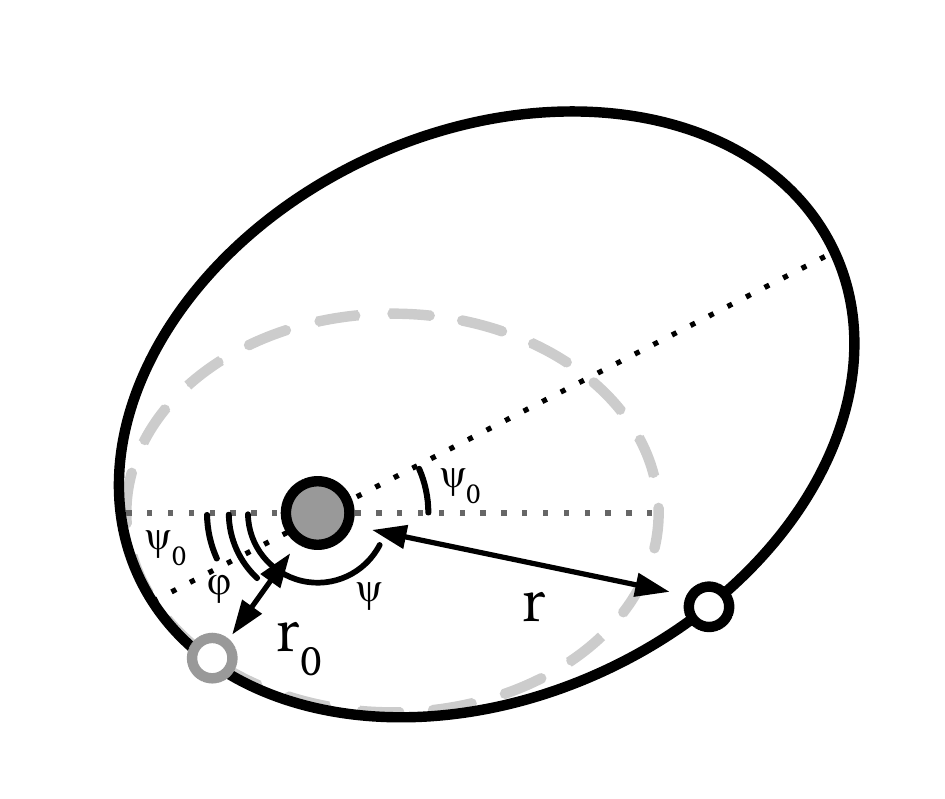}
\caption{Similar to Figure \ref{ellipse}. We show the initial orbit of
  a particle before mass loss (dashed grey) and the new orbit
  following mass loss (black). The position of the particle at the
  moment of mass loss is shown as a grey circle. $\phi$ is the phase
  of the particle at the moment of mass loss (with $\phi_0 = 0$, thus
  setting the orientation of the initial orbit). The new orbit is not
  in general aligned with the original and hence $\psi_0$, the
  difference in orientation between the two orbits, is not 0. The
  particle is shown at some arbitrary time, $t>0$, as a black circle,
  with some radius $r$ and phase $\psi$.}
\label{phaseEllipse}
\end{figure}

In \S~\ref{Kepler} we derived the response of a system of particles on
initially circular orbits to instantaneous mass loss.

The circular case is the simplest and most intuitive but far from the
only analytic case. Here we derive the orbital parameters after mass
loss for any initial Keplerian orbit.

The general results stated in \S~\ref{BasicKepler} are as useful here
but we'll also make use of a few more results.

The radial and tangential velocities can be found via
\begin{equation}
\label{velocities}
v_t=\frac{l}{r} \ \mathrm{and} \ v_r=\frac{le\sin(\phi - \phi_0)}{a(1-e^2)}.
\end{equation}

By finding the velocities at peri- and apoapsis (where the radial
velocity is 0) we find the Vis-Viva equation:
\begin{equation}
\label{vis-viva}
v^2=GM\left( \frac{2}{r} - \frac{1}{a} \right).
\end{equation}

\subsection{Response to mass loss}

Recalling equation \ref{kepler_basic} let us choose an initial configuration
\begin{equation}
\label{non_circ_basic}
r_0=r(t=0)=\frac{a (1-e^2)}{1+e \cos{\phi}}
\end{equation}
where we've set $\phi_0$ to be 0 (setting the orientation of our
co-ordinate system such that the particle passes through periapsis at
$\phi=0$) and $\phi$ is the orbital phase at the moment of mass loss.

Now the initial state of the system is expressed in 3 parameters,
$r_0,e$ and $\phi$, rather than a single parameter in the circular
orbit case, $r_0$. (Note that using $r_0$ is more convenient than $a$; however,
both suffice, and the conversion is trivial.) The problem is still
spherically symmetric, and we can still follow the evolution of shells
rather than individual particles, but now the shells correspond to
particles with the same $r_0, e$ and $\phi$. We have moved from a 1
dimensional parameter space to 3.

The new orbit will be of the similar form,
\begin{equation}
\label{kepler_new}
r=\frac{\alpha (1-\epsilon^2)}{1+\epsilon \cos(\psi - \psi_0)},
\end{equation}
where $\alpha$, $\epsilon$ and $\psi$ are the new semi-major axis,
eccentricity and phase respectively.

Figure \ref{phaseEllipse} shows an illustration of this for a single
particle. The initial and final orbits will have different
orientations, and this difference depends on the phase at the moment
of mass loss. Momentarily at $t=0$, $\psi=\phi$, i.e. both phases have
the same orientation; however, $\psi_0$ is not in general equal to 0.

The velocity the instant of the mass loss is unchanged, hence at $t=0$
the Vis-Viva equation (equation \ref{vis-viva}) is satisfied for both
the unperturbed and perturbed cases, with the same $v^2$. Equating the
RHS of both and rearranging gives
\begin{equation}
\label{ellipseAlpha}
\alpha=\frac{m}{M\left(\frac{1}{a}+\frac{2}{r_0} \left( 1-\frac{m}{M} \right) \right)}.
\end{equation}

The angular momentum of the orbit is constant throughout, and has the
same form in both the unperturbed and perturbed case, i.e. equation
\ref{kepler_l} with the relevant mass, eccentricity and semi-major
axis. Setting both expressions for $l$ equal and rearranging gives
\begin{equation}
\label{ellipseEpsilon}
\epsilon=\sqrt{1-\frac{M}{m}\frac{a}{\alpha}(1-e^2)}.
\end{equation}

The two expressions for radius must match at $t=0$, i.e. when
$\psi=\phi$, setting $r(\psi=\phi)=r_0(\phi)$ gives
\begin{equation}
\label{phi_0}
\psi_0=\phi - \cos^{-1}\left(\frac{1}{\epsilon} \left( \frac{M}{m} (1+e \cos (\phi) )\right) \right).
\end{equation}
There are two possible values of the arccos term, with a difference of
$\pi$, but the correct one can be chosen by ensuring
$\sin(\phi-\psi_0)$ and $\sin(\phi)$ have the same sign, i.e. the
direction of the radial velocity is consistent.

Thus for any combination of $(r_0,e,\phi)$, we can find the parameters
of the new orbit.

Setting $e=0$, we can easily recover the relations for circular orbits
given in \S~\ref{KeplerResponse}.

For the family of orbits with the same $e$ and $\phi$ we again find
that the new eccentricity is a constant and the semi-major axis is
linearly proportional to $r_0$. These orbits are again all similar,
differing only in period, and any profile will maintain its shape and
simply evolve in time as a re-scaling of the $r$ co-ordinate (using
$\frac{r}{t^{2/3}}=const.$).

Now the mass loss necessary for a particle to become unbound
($\epsilon>1$) depends on the initial phase and eccentricity. A
particle will be unbound for
\begin{equation}
\label{ecc_unbound}
\frac{m}{M} < 1-\frac{r_0}{2a} \ \left( =1-\frac{1-e^2}{2(1+e\cos{\phi})} \right).
\end{equation}
The left hand expression is smallest for particles initially at
periapsis, where $\phi=0$, hence for particles of a range of initial
phases at least some will be lost if $\frac{m}{M} < \frac{1+e}{2}$. If
there is also a distribution of initial eccentricities we may expect
it to include particles up to $e=1$ (but not including as these would
be unbound) and hence for any finite central mass loss some particles
must become unbound.

For particles of a given initial eccentricity and for a specific $m
\ (<\frac{1+e}{2})$ all those with
\begin{equation}
\label{phi_unbound}
\vert \phi \vert < \cos^{-1} \left( \frac{1}{e} \left( 1-\frac{2}{1-e^2} \left(1-\frac{m}{M} \right) \right) \right)
\end{equation}
will be unbound. This means that it is the particles closest to
periapsis that are easiest to lose from the system.

The radius, and the time since mass loss, can still be expressed
simply via equations \ref{kepler_r} and \ref{kepler_t}, although now
$t_0$ will not in general be 0 (as in general particles do not start
at periapsis). We can find $t_0$ using equation \ref{eta} using
$e=\epsilon$, $\phi_0=\psi_0$ and the phase at the moment of mass
loss, $\phi$. Hence the same techniques used in \S~\ref{KeplerDensity}
(and further detailed in Appendix \ref{KeplerSolve}) could be used to
find the density profile.

To do this for shells with a continuous range of $r_0, e$ and $\phi$
would, however, require a root-finding in 3 dimensions and is
significantly more computationally complex.

\section{Computing the density profile}
\label{KeplerSolve}

\begin{figure}
\includegraphics[width=\columnwidth]{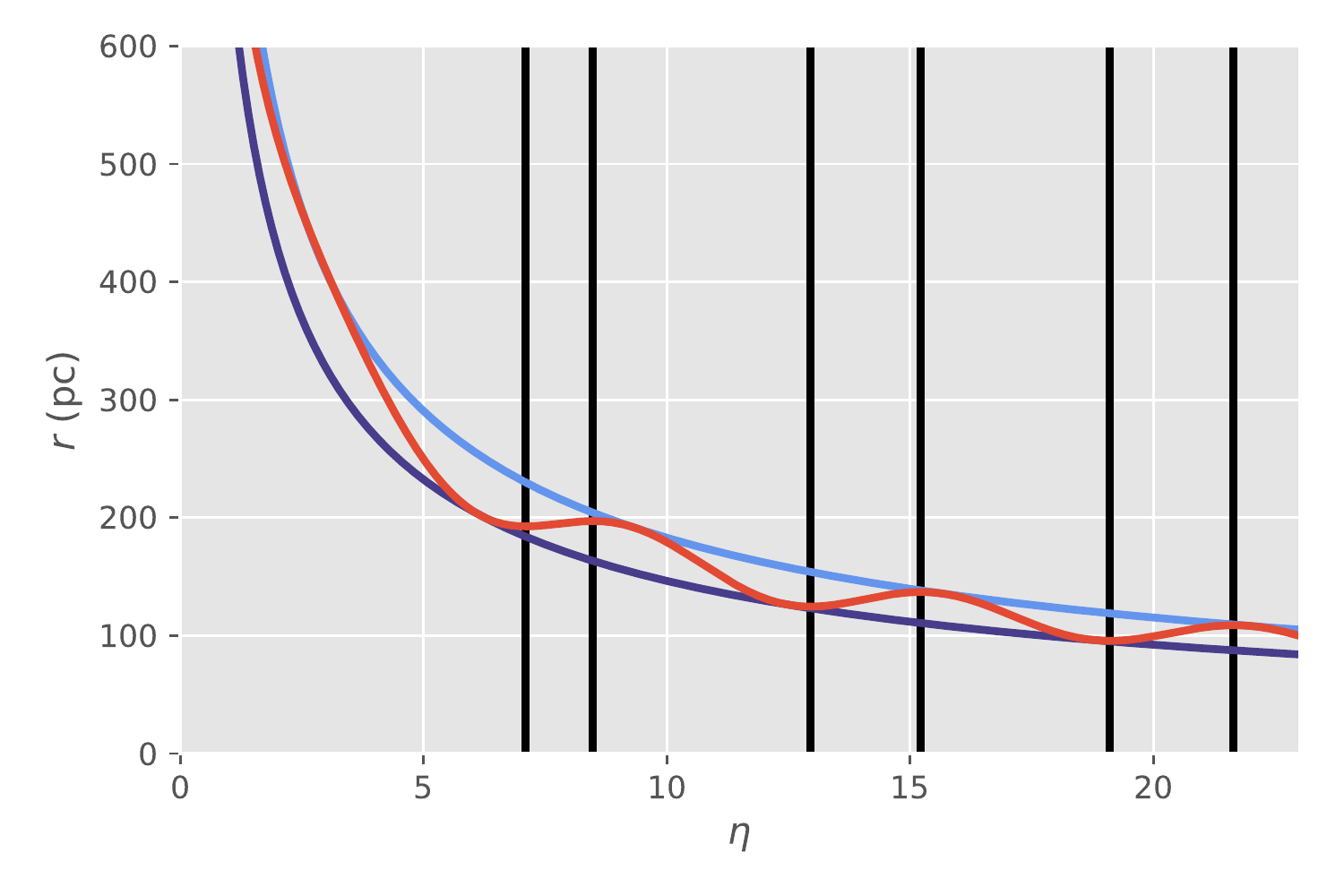}
\caption{Plotted is equation \ref{new_ri} (red) and the bounding
  curves given in equation \ref{plusminus_r} (light and dark
  blue). The turning points of the function are also plotted
  (black). The $t$, $M$ and $m$ parameters are the same as Figure
  \ref{r_r0}.}
\label{r_eta}
\end{figure}

\begin{figure}
\includegraphics[width=\columnwidth]{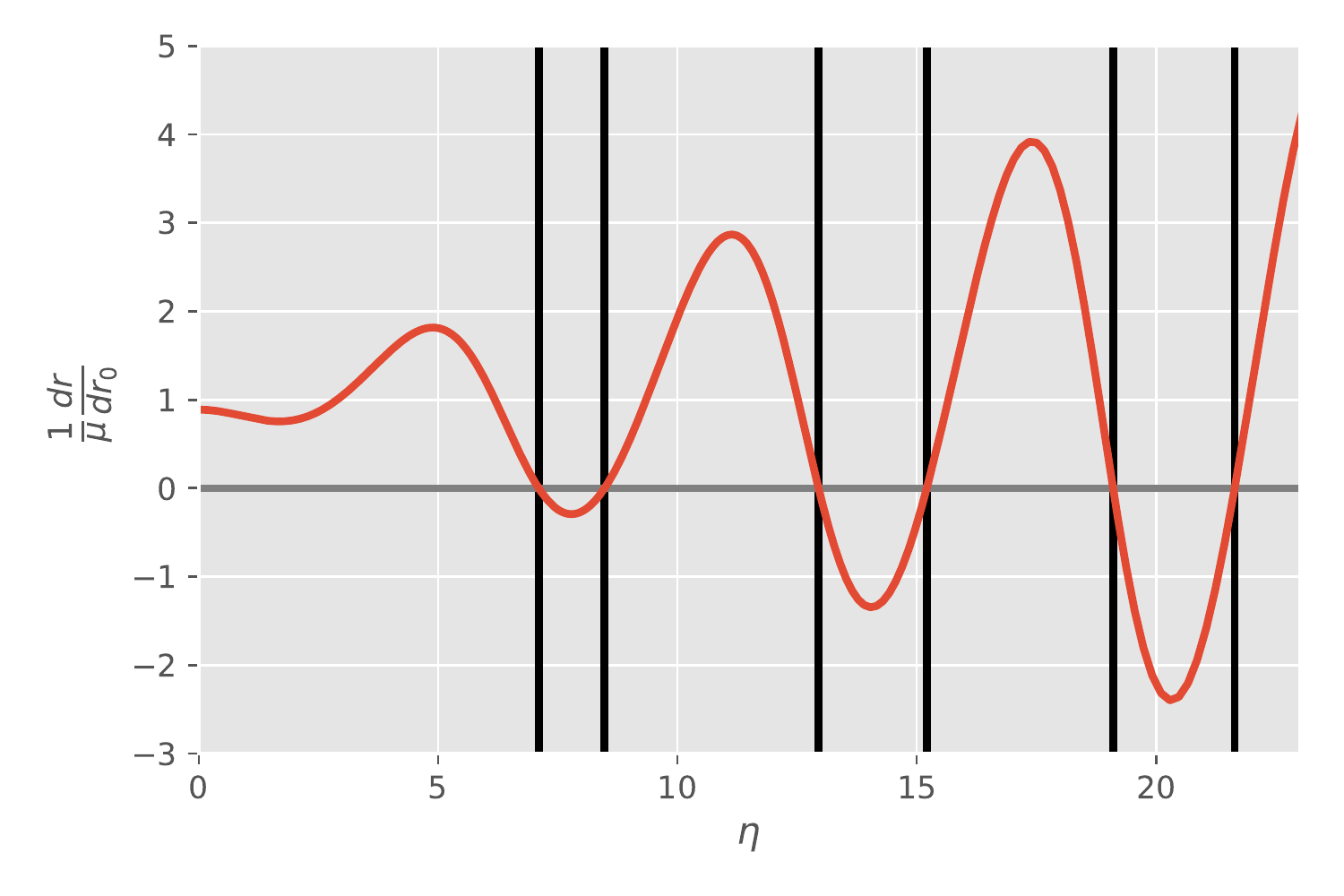}
\caption{The gradient of Figure \ref{r_r0}, normalised by $\mu$, from
  equation \ref{dr_dr0} plotted against $\eta$ (red). The roots of the
  function are shown (black). $M$ and $m$ are the same as Figure
  \ref{r_r0}.}
\label{omega_eta}
\end{figure}

In \S~\ref{KeplerResponse} we derived a simple form for the variations
in radius of a shell, initially on a circular orbit, that could be
solved numerically for a given $t$ and $r_0$. Here we show how to use
these equations, and variations thereof, to find the density profile
via a 1D root finding of a well-behaved function.

A similar analysis could be used for the non-circular case, using the
results from Appendix \ref{EllipseKepler}, for a given $t, r_0,
\epsilon$ and $\phi$. It is, however, substantially more convoluted
and ultimately unnecessary within the scope of this paper, so we will
only delve into the circular case.

To find which shells are currently at a given radius we can use
equation \ref{new_reta}, which we'll rewrite as the radius of an
individual shell,
\begin{equation}
\label{new_ri}
r_i(t,\eta_i)=\left( G m t^2 \frac{(1-\epsilon \cos{\eta_i})^3}{(\eta_i - \epsilon \sin{\eta_i})^2} \right)^\frac{1}{3},
\end{equation}
given its corresponding $\eta_i$.

This curve is bounded by
\begin{equation}
\label{plusminus_r}
r_{\pm}(t,\eta)=\left( G m t^2 \frac{(1 \pm \epsilon)^3}{\eta^2} \right)^\frac{1}{3}.
\end{equation}

All three curves are shown in Figure \ref{r_eta}.

The turning points of equation \ref{new_ri} are the same as those of
equation \ref{dr_dr0} (from
$\frac{dr}{dr_0}=\frac{dr}{d\eta}\frac{d\eta}{dr_0}$ and using
$\frac{dr_0}{d\eta}>0$ for all $\eta$).

Figure \ref{omega_eta} shows $\frac{1}{\mu}\frac{dr}{dr_0}$ and its
roots. These roots must be found numerically. This is relatively easy
given that the curve has clear periodic behaviour and hence the
$n^{th}$ root, $\eta_{r,n}$, must lie between $(2k+n)\pi$ and
$(2k+n+1)\pi$, where $k$ is an integer, dependent on $\epsilon$, which
sets the offset of the first root.

For the case shown in Figure \ref{omega_eta}, the value of $k$ is
clearly 1. For a larger mass loss, and hence a larger $\epsilon$, it is
possible for $k$ to equal 0, and for smaller mass loss, the first root
may be at much higher $\eta$. Given that $r$ is inversely proportional
to $\eta$, physically this corresponds to the furthest caustic being at
a larger radius. These roots are independent of time and hence need
only be calculated once, and though there is an infinite number as $n$
increases, the $n^{th}$ root soon corresponds to vanishingly small
radii.

Putting this in simpler terms, there are an infinite number of
caustics going down to $r=0$, and dependent on the fractional mass
loss, $\frac{m}{M}$, the first caustic can correspond to different
values of $\eta$. For minima in Figure \ref{omega_eta} with
$\frac{dr}{dr_0} > 0$ (if any exist), this corresponds to a smooth
bump, rather than a singularity, in the density profile.

With the turning points and the bounds in hand, we can now find all
$\eta_i$ for which $r_i=r$, i.e. all values of $\eta$ intesecting a
horizontal line in Figure \ref{r_eta}.

Rearranging equation \ref{plusminus_r} we can find, for a given $r$,
the maximum and minimum possible values of $\eta$,
\begin{equation}
\label{plusminus_eta}
\eta_{\pm}(t,r)=\sqrt{\frac{GMt^2}{r^3}(1 \pm \epsilon)^3}.
\end{equation}

Thus there is some $\eta_i$ for which $r_i(\eta_i)-r=0$ for each
interval from $\eta_{r,m}$ to $\eta_{r,m+1}$ where $m$ runs from the
smallest $n$ such that $\eta_{r,n} > \eta_-$ (inclusive) to the
largest $n$ such that $\eta_{r,n} < \eta_+$ (exclusive). It is
computationally easy to find each of these roots independently. There
may also be roots in the two immediately adjacent intervals, but this
is dependent on $r$ and these must be checked independently.

With the full range of $\eta_i$ we can then, finally, enumerate
equation \ref{rho} (using equations \ref{new_r} and \ref{dr_dr0}) and
hence find the density at any given $r$.

As all the functions we have had to explore numerically are
well-behaved, the corresponding density profile is truly analytically
correct, up to the limits of numerical machine precision.

\section{Perturbation analysis of caustics}
\label{Perturbation}

Here we reproduce the behaviour of the density profile as it
approaches a caustic by looking at how the density profile changes
with radius very close to the singularity. This intended as an
analytic derivation of the numerical results from Figure \ref{n_r},
where we find that the profile approaches the singularity as an approximate
power law with exponent of $-\frac{1}{2}$.

Taking a single shell with density $\rho_i$, we can expand the square
of the density,
\begin{equation}
\label{rho_i}
\rho_i^2=\left( \frac{r_0}{r} \right)^4 \left(\frac{dr_0}{dr} \right)^2 \rho_0(r_0)^2,
\end{equation}
(here we use the square to save worrying about the absolute value). As
$\frac{dr}{dr_0}$ is easiest to express in $\eta$, we will expand
around $\eta_0$, the value of $\eta$ corresponding to the caustic with
$\left( \frac{dr}{dr_0} \right)_{\eta_0} = 0$.

We evaluate equation \ref{rho_i} at some
\begin{equation}
\label{new_eta}
\eta=\eta_0 + \Delta\eta.
\end{equation}
We can convert this to the variation around the initial radius of the
shells corresponding to this caustic,
\begin{equation}
\label{diff_r0}
\Delta r_0 = \frac{dr_0}{d\eta} \Delta \eta + O(\Delta \eta^2) = -\frac{2 r_0 (1-\epsilon \cos{\eta})}{3(\eta-\epsilon \sin{\eta})} \Delta \eta + O(\Delta \eta^2).
\end{equation}

Near the caustic we can Taylor expand $\frac{dr}{dr_0}$ to give
\begin{equation}
\label{new_grad}
\left( \frac{dr}{dr_0} \right)_{\eta} = \left( \frac{dr}{dr_0} \right)_{\eta_0} + \left( \frac{d}{d\eta} \left( \frac{dr}{dr_0} \right) \right) \Delta \eta + O(\Delta\eta^2)
\end{equation}
but at the caustic the first term disappears and only the second term
is left, giving
\begin{equation}
\begin{split}
\label{eurgh}
\left( \frac{dr}{dr_0} \right)_{\eta} &= -\frac{\epsilon}{2} \left( \sin{\eta} \right. \\ &\left. + 3\left(\frac{\eta -\epsilon \sin{\eta}}{1-\epsilon \cos{\eta}} \right) \left(
\cos{\eta} - \frac{\epsilon \sin^2\eta}{1-\epsilon \cos{\eta}}  \right) \right)_{\eta_0} \Delta \eta + O(\Delta\eta^2)
\end{split}
\end{equation}
where this form is given only to show that the co-efficient of the
$\Delta\eta$ term is non-zero at the caustics.

As $\frac{dr}{dr_0}$ contains only terms linear in $\Delta \eta$ or
higher, and $\frac{r}{r_0}$ does not go to zero at the caustics
\begin{equation}
\label{denom}
\left( \frac{r_0}{r} \right)^4 \left(\frac{dr_0}{dr} \right)^2 \propto \Delta \eta^{-2} (1+O(\Delta \eta^2))^{-1}.
\end{equation}

We could expand the initial density using equation \ref{diff_r0} but
it is simple to show that any terms beyond the $\rho_0(r_0)$ term are
negligible.

Hence
\begin{equation}
\label{rho_sq_eta}
\rho_i^2 \propto \Delta \eta^{-2} (1+O(\Delta \eta^2))^{-1}
\end{equation}
or
\begin{equation}
\label{rho_sq_r0}
\rho_i^2 \propto \Delta r_0^{-2} (1+O(\Delta r_0^2))^{-1}.
\end{equation}

Finally we can translate this to variation in $r$, $\Delta r$, where
\begin{equation}
\label{delta_r}
\Delta r = \left( \frac{dr}{dr_0} \right)_{r_c} \Delta r_0 + \left( \left( \frac{d}{dr_0} \frac{dr}{dr_0} \right) \right)_{r_c} \Delta r_0^2 + O(\Delta r_0^3) 
\end{equation}
where again the first term on the right goes to zero (and it is simple
to show the second term is non-zero at $r_c$). Thus $\Delta r \propto
\Delta r_0^2 + O(\Delta r_0^3)$.

Putting this back in to equation \ref{rho_sq_eta} and taking the
square root finally yields
\begin{equation}
\label{rho_delta_r}
\rho_i \propto \Delta r^\frac{1}{2}
\end{equation}
to lowest order. As expected, this fits with Figure \ref{n_r} as we
approach the location of the caustic. This analysis also is completely
general to any caustic, regardless of whether it approaches the
singularity from above or below.

\section{The \textsc{CausticFrog} package}
\label{Code}

\begin{figure}
\includegraphics[width=\columnwidth]{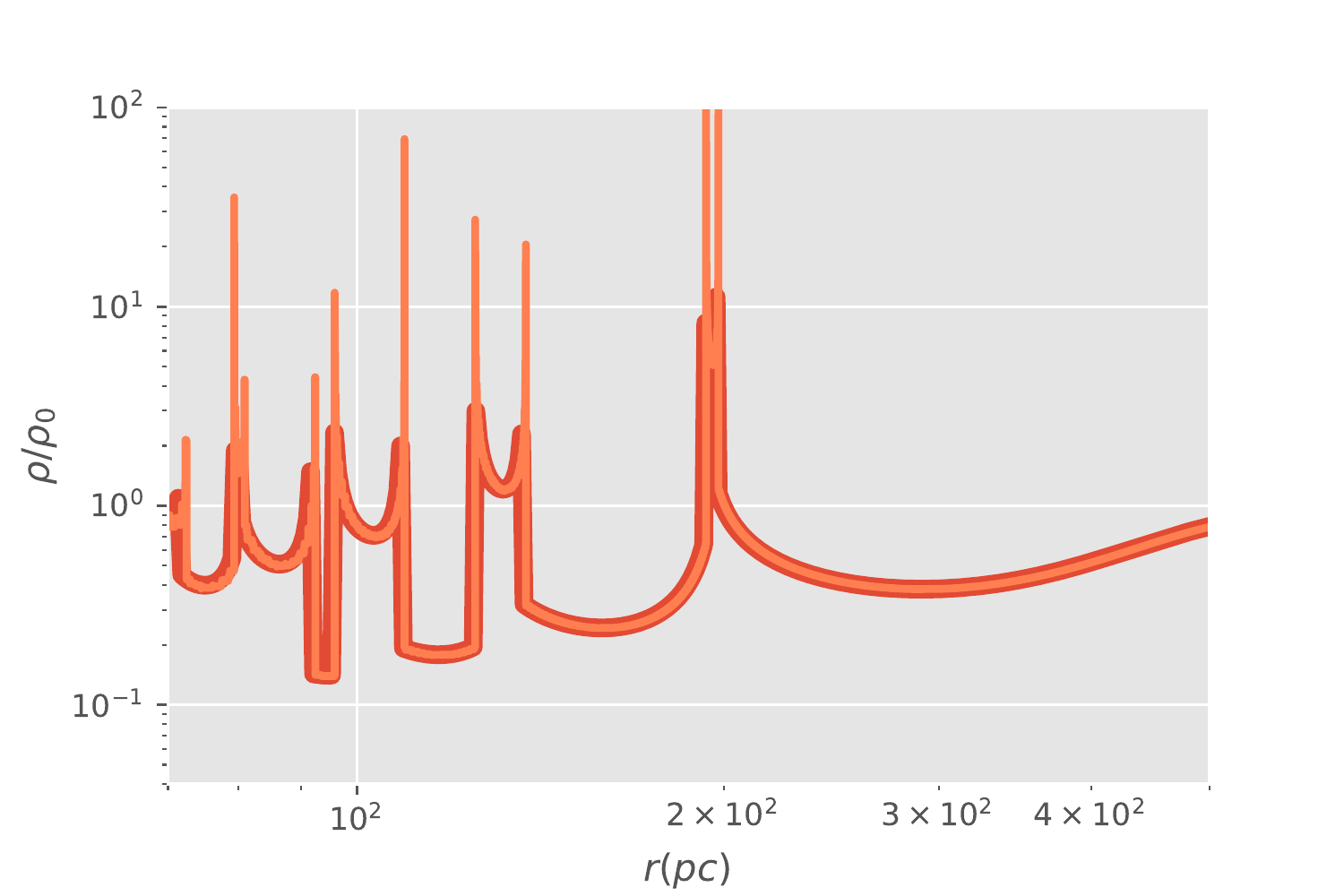}
\caption{The analytic density profile for initially circular orbits in
  a Keplerian potential (thick red) compared to the same profile
  reproduced using \textsc{CausticFrog} (thin orange).}
\label{sim_rho}
\end{figure}

Many previous studies of systems undergoing rapid mass loss have
modelled the evolution using test-particles, standard N-body or
hydrodynamics codes \citep{Lippai08,Shields08}. However, these will
fail to capture interesting features of these systems, such as the
squeezing of finite mass into negligible volume which causes the
caustics.

Instead, we present a new code, \textsc{CausticFrog}, designed
specifically to resolve this behaviour, while also exploiting the
symmetries of the system to simplify computation.

As shown in \S~\ref{Kepler}, the system will always remain spherically
(or, in the case of a disk, cylindrically) symmetric and we need not
model the motion of individual particles, but can follow the simpler
evolution of spherical shells. Our code is effectively Lagrangian,
following the evolution of fixed mass shells as they move and stretch
radially.

To simplify the terminology we'll use throughout this section, a
\textit{shell} contains a fixed finite amount of mass, $m_i$, and is
bounded by two \textit{edges} whose radii, $r_{a,i}$ and $r_{b,i}$, we
evolve directly. While each shell has a fixed mass enclosed, many
shells can overlap, leading to the density at that point being the sum
of the densities of all those shells.

We use a leapfrog integration, where at each moment in time the mass
enclosed by an edge is calculated as
\begin{equation}
\begin{split}
\label{massEnclosed}
M_{enc,i} =& m_{enc}(r_i) +\sum_{r_{a,j}, r_{b,j} < r_i} m_j + \sum_{r{a,j} < r_i < r{b,j}} m_j \frac{r_i^3 - r_{a,j}^3}{r_{b,j}^3 - r_{a,j}^3} \\
&+ \sum_{r{b,j} < r_i < r{a,j}} m_j \frac{r_i^3 - r_{b,j}^3}{r_{a,j}^3 - r_{b,j}^3}
\end{split}
\end{equation}
where $m_{enc}$ includes any mass enclosed that is not part of the
gravitating fluid (e.g. for the Keplerian potential this would be the
central point mass).

The resulting acceleration on the shell therefore is
\begin{equation}
\label{acceleration}
\ddot{r}_i = \frac{1}{r_i^3} (l_i^2 - G M_{enc,i} r_i)
\end{equation}
where $l_i$ is the specific angular momentum of the edge and is
constant throughout (as there are no tangential impulses).

An edge with a given initial radius, eccentricity, and phase
($r_{0,i},e_i,\phi_i$) is initialised via equation \ref{kepler_l} for
the angular momentum $l_i$ and equation \ref{velocities} for the
initial radial velocity $v_{r,i,0}$.

Rather than evolve many separate shells we follow the evolution of a
"accordion" of shells, where two consecutive shells share the same
edge, i.e. $r_{b,i}=r_{a,i+1}$. Each accordion has a single initial
eccentricity and phase. Grouping shells in this way halves the
computation time (as now there is effectively one unique edge per
shell).

In this paper, we only show results from simulations in Keplerian
potentials or variations thereof, but the code can accept any mass
profile for the gravitating particles, and for any external mass
before and after mass loss.

The code is written in \textsc{Python} and \textsc{Cython}, and can be
found on \textsc{GitHub} at
\url{https://github.com/zpenoyre/CausticFrog}. There is an example
\textsc{iPython} notebook showing how to initialise and run
simulations.

As a simple code test, Figure \ref{sim_rho} shows the density profile
recovered for a Keplerian potential, compared to the analytic solution
shown in \S~\ref{Kepler}.

We encourage anyone who wishes to use the code to contact us so we can
provide advice and assistance.


\bsp	
\label{lastpage}
\end{document}